\documentclass[aps,prd,amsmath,amssymb,twocolumn,superscriptaddress,preprintnumbers,nofootinbib,floatfix]{revtex4-1}
\pdfoutput=1
\usepackage{graphicx}
\usepackage{dcolumn}
\usepackage{bm}
\usepackage{textcomp}
\usepackage{aas_macros}

\newcommand{\be}{\begin{eqnarray}}
\newcommand{\ee}{\end{eqnarray}}
\newcommand{\bea}{\begin{eqnarray}}
\newcommand{\eea}{\end{eqnarray}}

\newcommand{\ER}{E_{_R}}
\newcommand{\vratio}{\epsilon_v}
\newcommand{\keVee}{~{\rm keV_{ee}}}
\newcommand{\vmin}{v_{\rm min}}

\begin{document}

\title{Statistical Tests of Noise and Harmony in Dark Matter Modulation Signals}
\author{Spencer Chang}
\email{chang2@uoregon.edu}
\affiliation{Department of Physics, University of Oregon, Eugene, OR 97403}
\author{Josef Pradler}
\email{jpradler@perimeterinstitute.ca}
\affiliation{Perimeter Institute for Theoretical Physics 31 Caroline St. N, Waterloo, Ontario, Canada N2L 2Y5.}
\author{Itay Yavin}
\email{iyavin@perimeterinstitute.ca}
\affiliation{Perimeter Institute for Theoretical Physics 31 Caroline St. N, Waterloo, Ontario, Canada N2L 2Y5.}
\affiliation{Department of Physics \& Astronomy, McMaster University
1280 Main St. W, Hamilton, Ontario, Canada, L8S 4L8}


\begin{abstract}
The aim of the current work is a detailed time-series analysis of the
data from Dark Matter direct detection experiments as well as related
datasets. We examine recent claims that the cosmic ray muon flux can
be responsible for generating the modulation signals seen by DAMA and,
more recently, by the CoGeNT collaboration. We find no evidence for
such a strong correlation and show that the two phenomena differ in
their power spectrum, phase, and possibly in amplitude.  In addition,
we investigate in more detail, the time dependence of Dark Matter
scattering.  Since the signal is periodic with period of a year (due
to the Earth's motion around the Sun), the presence of higher
harmonics can be expected.  We show that the higher harmonics
generically have similar phase to the annual modulation and the
biannual mode in particular could provide another handle in searching
for Dark Matter in the laboratory.
\end{abstract}

\maketitle

\section{Introduction}
Whereas the gravitational nature of Dark Matter (DM) is a crucial
ingredient for the success of the standard cosmological model, its
non-gravitational character proves elusive and remains essentially
unknown to date. The possible connection with the electroweak scale
and the associated expectation of an interaction strength not much
smaller than $G_F$ and a DM mass not much lighter than
$\mathcal{O}(10\,\mathrm{GeV})$ has motivated the construction of
experiments looking for the direct scattering of DM against nuclei in
the lab~\cite{witten}. The experimental efforts in the field are
proceeding in full force and at least another decade of progress is
expected.

With the main observable being the nuclear recoil spectrum the
information content is rather limited. Moreover, as was realized over
the past decade, the recoil spectrum resulting from DM collisions is
model dependent and may not follow the simple exponential rise towards
low energies expected from elastic scattering~\cite{iDM,CiDM, MDDM, MiDM,
  FFDM}. On the other hand, one of the most robust predictions of the
cold DM model is that the relative velocity of the Earth against the
DM halo should vary with the Earth revolution around the
Sun~\cite{Drukier:1986tm, Freese}. The expected recoil rate is therefore a general
periodic function with a fundamental period of one year with a
particular phase. Despite the strong motivations to look for such
modulations, it has so far been achieved by only two experiments,
DAMA~\cite{Bernabei:2008yi,Bernabei:2010mq} and
CoGeNT~\cite{Aalseth:2011wp}. The difficulty is of course in
maintaining the stability of the apparatus and the rejection of
background over a long period of time. Indeed, the two experiments
just mentioned have experienced considerably more background events
than some of the other extremely clean experiments such as
\mbox{CDMS-II}~\cite{cdms,Ahmed:2010wy} and
XENON100~\cite{xenon100}. Moreover, the search for annual modulations
is not without difficulties since many more mundane phenomena are
known to exhibit such modulations. Therefore, two of the main goals of
this paper are 1) to show that the time series of the reported DAMA
and CoGeNT signals can be used to make firm statements about
background hypotheses proposed in the literature and 2) to establish
further observables based on the time distribution of the energy
spectrum.

The only direct detection experiment which claims detection of a firm
signal from dark matter is DAMA, situated in the underground LNGS
laboratory at Gran Sasso, Italy. The DAMA dataset consists of two main
periods, DAMA/NaI (Dec 1995 - July 2002) and DAMA/LIBRA (Sept 2003 -
Sept 2009), amounting to a cumulative exposure of
1.17~ton$\times$years. The residual rate shown by the collaboration
exhibits a very clear annual modulation compatible with what is
expected from DM models where the Earth's motion around the Sun
results in a modulation peaking on approximately June 2$^{\rm
  nd}$. The collaboration has also released the modulation amplitude
as a function of recoil energy, which seems to exhibit a peculiar form
possibly more consistent with inelastic scattering~\cite{iDM,CiDM,MDDM,
  FFDM,MiDM}. It should be kept in mind, however, that this energy
spectrum is obtained from the full data set by assuming the periodic
functional form. Sadly, there has been no release of the full
data set to date. It is particularly unfortunate because, as we will
see below, the procedure carried by the collaboration to obtain the
power spectrum of the signal and other parameters makes it difficult
to compare to both background and signal hypotheses.

Given the seasonal nature of the signal it is entirely conceivable
that environmental effects induce backgrounds with an annual
modulation pattern just like the one seen by DAMA. As such, it is
clear that such contamination may depart significantly from a
sinusoidal distribution in time. It is therefore important---whenever
enough data on a putative background inducing process is
available---to assess its viability based on a full time-series
analysis. For example, in this work we employ Pearson's coefficient of
correlation as a test statistic when quantifying the compatibility of
two data sets. This allows for a quantitative comparison between two
data sets which is independent of any assumption about the functional
form of the time variations.

Fluxes of neutrons probably constitute the most dangerous source of
background as they lead to nuclear recoils which can mimic DM-nucleus
interactions \cite{ralston}. Indeed, it is known from ICARUS
measurements at LNGS \cite{icarus} that the ambient neutron flux
generated in the surrounding rock shows some variation on the
timescale of a year. However, with a total of five data points we find
that it is not possible to make any \textit{statistically} significant
statement regarding the temporal compatibility of this candidate
background with the DAMA signal. We therefore choose not to elaborate
on this issue any further.

\begin{figure}[tb]
\begin{center}
\includegraphics[width=\columnwidth]{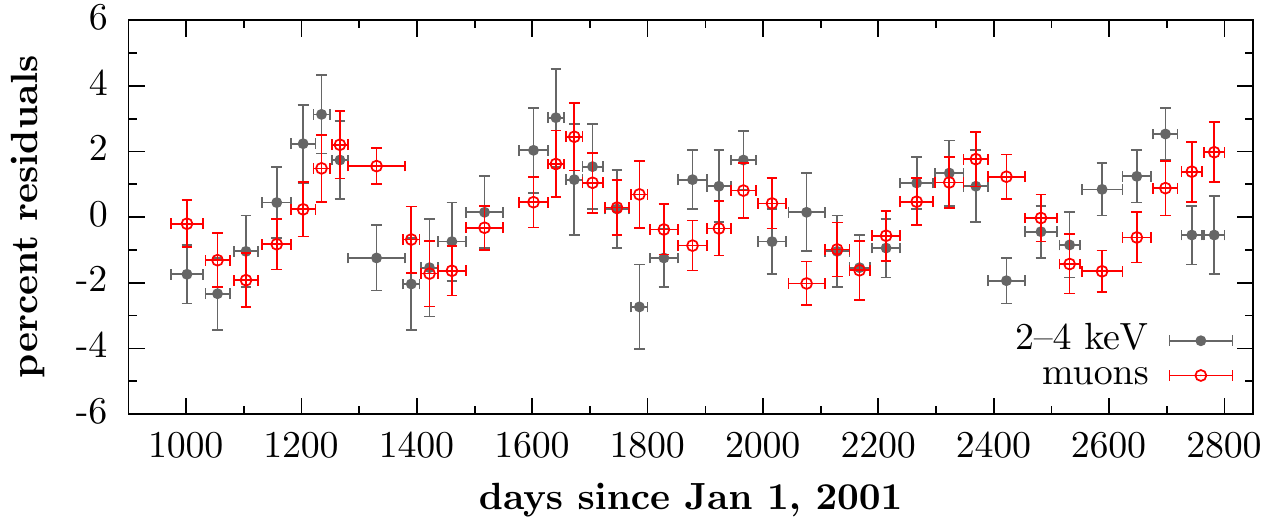}
\end{center}
\caption{Percent annual residuals of the LVD measured muon flux when
  binned in accordance with the DAMA/LIBRA runs~1--5. The latter
  residuals are shown for the 2--4\,keV bins assuming a baseline count
  rate of $\bar s = 1.15$~cpd/kg/keV.}
\label{fig:lngs}
\end{figure}

In contrast to rock-generated neutrons, a wealth of data is available
on another guaranteed source of neutrons: cosmic ray muons which can
penetrate deep underground and induce spallation reactions in the
detector and nearby. It is also possible for these muons to deposit
their energy directly into the detector crystals~\cite{nygren}.
Measurements of the muon flux underground have a long history and its
seasonal variation is firmly established. There are published
measurements at LNGS from MACRO~\cite{Ambrosio:1997tc},
LVD~\cite{Selvi} and most recently from
Borexino~\cite{D'Angelo:2011fs}. For DAMA, the relevant measurement is
the one from LVD since it was taken at the same underground lab and it
overlaps in time with the DAMA/LIBRA runs 1--5. Figure~\ref{fig:lngs}
shows the percent residuals of the muon flux when binned in
concordance with DAMA with the annual mean count rate subtracted. Also
shown are the residuals in the 2--4\,keV bin of the DAMA/LIBRA data
assuming a baseline rate of $\bar s = 1.15$~cpd/kg/keV. The seeming
similarity in time and amplitude is tantalizing and it has lead
various authors to put forward the hypothesis that both signals are in
fact measurements of the very same cosmic ray
phenomenon~\cite{ralston,nygren,Blum:2011jf}. It will be one of the
central points of the paper to critically examine these claims,
finding that they have difficulty explaining the observed modulations.

The issue of annual modulation in direct DM detection has recently
received further impetus from the CoGeNT experiment, located in the
Soudan Mine, MN, USA. The collaboration has published its analysis on
a 458 day run with a 440\,gram Ge detector
\cite{Aalseth:2011wp}. After removal of cosmogenic background
contamination, the data exhibits a seasonal variation peaking around
the middle of April. As we shall see below the modulation is manifest
only in the higher energy part of the recoil spectrum where the energy
spectrum is rather flat with none of the usual features expected from
the direct detection signal of DM, elastic or otherwise. The
unexplained exponential rise in the recoil spectrum at lower energies,
which has stirred quite a commotion in the recent past, shows no
evidence of annual modulations. Making use of the time-stamped data
provided by the collaboration we address the modulation part of the
signal (see also ref.~\cite{Fox:2011px,volansky} for prior analyses)
and investigate the potential role of cosmic ray muons for the CoGeNT
detector.

In light of the significant advances in sensitivity from direct
detection experiments achieved in the past decade and future
improvements expected in the next one, we also address the question of
whether the time-series of a signal may encode additional evidence
that it is DM which is scattering on a target. In particular we point
out that higher harmonics may be present in an annually modulated
signal. We show how this signature manifests itself in the scattering
rate and explore to some extent its sensitivity on the assumed halo
parameters.

The paper is organized as follows: We start with
section~\ref{sec:null} by establishing statistical evidence for
periodic variation in the data sets under consideration. Restricting
ourselves to a sinusoidal form of the signals, in
section~\ref{sec:phase} we allow both phase and period to vary and
explore the inferred values of these two parameters from the different
datasets.  In section~\ref{sec:correlation} we drop the assumption of
a sinusoidal form and instead directly compare the time series of the
various data sets by performing a correlation
analysis. Section~\ref{sec:biannual} discusses the effect of higher
harmonic modulations as a new diagnostic tool when searching for dark
matter in the lab.  We summarize the main conclusions and findings of
the paper in section~\ref{sec:conclusions}.

\section{Rejecting the Null Hypothesis}
\label{sec:null}
The existence of a periodic signal in the DAMA and LVD data sets is
clear even without any sophisticated statistical analysis. The
situation is somewhat more subtle in the case of CoGeNT. Both for the
purpose of completeness, and because it reveals interesting
differences between the data sets, we begin our analysis by testing
the null hypothesis of no signal (\textit{i.e.}~only pure noise) in
each of the data sets. The most convenient way of doing such a
spectral analysis is by looking at a periodogram of the data.

The classical periodogram of Schuster~\cite{schuster1890} allows for a
calculation of the power spectrum of a signal even in the case of
uneven sampling of the data. In ref.~\cite{1982ApJ...263..835S}
Scargle showed that a modified version of the classical periodogram,
which we call the Lomb-Scargle (LS) power spectrum (and sometimes just
power spectrum), results in the same well defined statistical behavior
as the Fourier power spectrum used in the case of even sampling. The
LS power spectrum allows one to reject noise at a $1-\alpha$
confidence level for a single, preselected frequency by demanding a
power level in excess of $z_\alpha = \ln\left(\alpha^{-1}\right)$. If
one is testing $N$ independent frequencies then the power level
required increases to $z_\alpha =-\ln \left[ 1-(1-\alpha)^{1/N}
\right]$ which contains the statistical penalty for inspecting more
than one frequency. This means that when testing for the null
hypothesis over a range of frequencies we must employ some care in
choosing the frequencies to be tested if we want a meaningful
statistical interpretation. If the time series is evenly distributed
then the standard choice for the frequencies in the discrete Fourier
transform results in independent frequencies. Since the data sets we
consider are not grossly uneven and approximately cover an integer
number of years we chose $\omega_n = n \omega_F$ with the fundamental
frequency $\omega_F = 2\pi/T$ where $T$ is roughly the range of years
covered by the data set\footnote{The power spectra in this section all
  display a peak at a period of exactly one year. We caution the
  reader that this does not imply the best fit value would be exactly
  one year, but is simply an artifact of the frequencies we chose to
  sample when looking to reject the null hypothesis.}. We checked that this results in low correlation among the
test frequencies using the procedure described in App.~D of
Ref.~\cite{1982ApJ...263..835S}.

\subsection{DAMA and LVD data}

We start by considering the measurements of the muon flux by the LVD
experiment~\cite{Selvi}. The average integral muon intensity
underground is reported as $\langle I_{\mu} \rangle \simeq 3\times
10^{-4}\,\mathrm{m}^{-2}\,\sec^{-1}$ with an annual variation of $\sim
2\%$ in amplitude as can be seen from Fig.~\ref{fig:lngs}. The data
spans a total of eight years with more than 2800 data points which we
obtain by digitizing Fig.~2 of~\cite{Selvi}.

\begin{figure}[t]
\begin{center}
\includegraphics[width=\columnwidth]{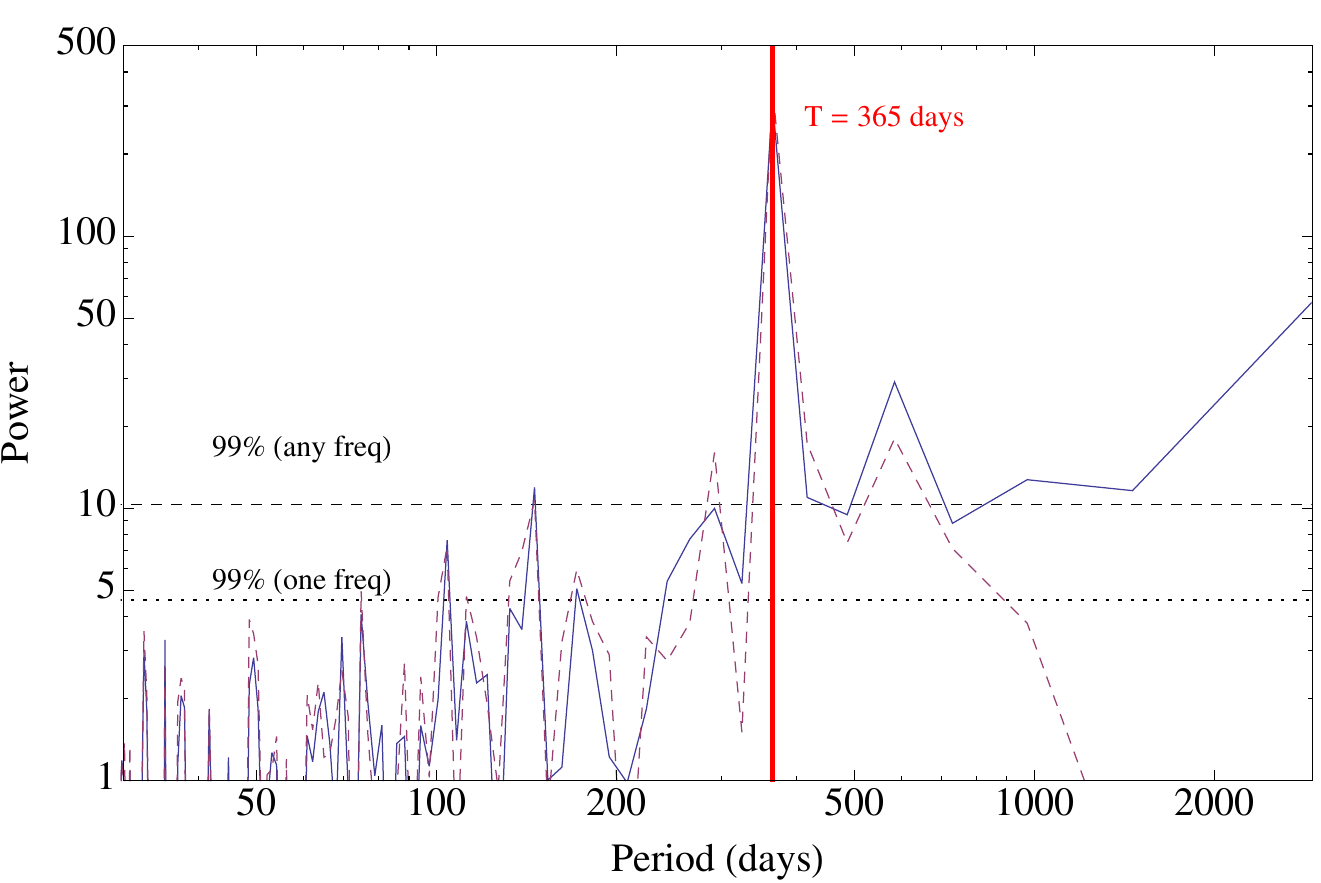}
\end{center}
\caption{The Lomb-Scargle power spectrum of the LVD data as a function
  of the period (solid-blue). Also drawn are the 99\% confidence lines
  for excluding the noise hypothesis for a single frequency (lower) or
  any of the frequencies shown (top). Note the substantial power
  present for modes with a period greater than a year. The dashed-red
  curve shows the effect of subtracting the yearly mean from the data
  on a yearly basis, as is done by the DAMA collaboration. The
  subtraction has little effect on the high frequency modes, but
  results in a strong damping of the long period modes as can be
  expected.}
\label{fig:LVD_null}
\end{figure}

The solid line in Fig.~\ref{fig:LVD_null} shows the power spectrum
obtained from the full LVD data set. Aside from the yearly modulation,
which is plainly visible from the time series itself, the power
spectrum also speaks unequivocally of the existence of temporal
variations on time-scales greater than a year. That same figure
(dashed line) shows the effect of subtracting the mean intensity from
the data on a \textit{yearly} basis as done by the DAMA collaboration
with their own data. The figure makes it clear that such a procedure
would mask most of the power at periods much greater than a
year. Nevertheless, we note that substantial power remains in modes
with periods a little over one year.

\begin{figure}[t]
\begin{center}
\includegraphics[width=\columnwidth]{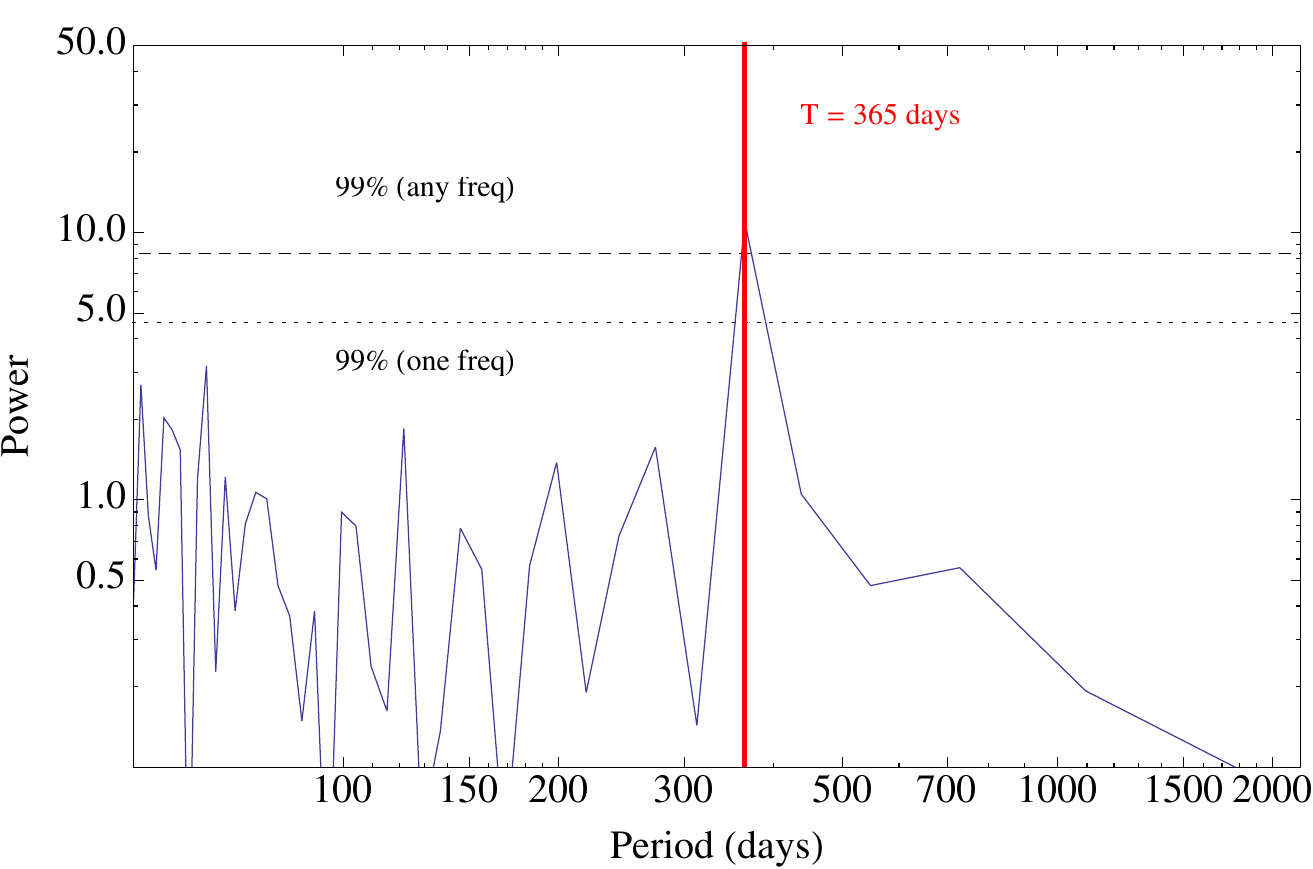}
\end{center}
\caption{The Lomb-Scargle power spectrum of the DAMA-LIBRA data in the
  $2-4\keVee$ energy region as a function of the period (solid-blue).}
\label{fig:DAMA-LIBRA_null}
\end{figure}

Figure~\ref{fig:DAMA-LIBRA_null} shows the power spectrum for the
DAMA/LIBRA $2-4\keVee$ data set. In contrast to the LVD spectrum,
there is little power at periods greater than a year. This, however,
may simply be an artifact of the way the DAMA/LIBRA modulation data is
obtained. The DAMA collaboration calculates the residuals by
subtracting the mean on a yearly basis for each cycle. As mentioned in
the previous paragraph such a procedure tends to dampen power at
periods much greater than a year. Nevertheless, the absence of any
power even for periods slightly over a year, as seen in the LVD data
above, is interesting and serves as the first distinguishing feature
between the two data sets. Unfortunately, it is difficult to assign a
quantitative significance to this difference without the full,
unsubtracted time series. This issue further motivates the release of
the unsubtracted data by the DAMA collaboration to allow for a proper
comparison of the power spectrum. Finally, we note that no biannual
mode (T =  1/2 year) is present in the DAMA/LIBRA power spectrum. We
will comment on this further in Sec.~\ref{sec:biannual} below.

\subsection{CoGeNT data}

The CoGeNT collaboration has released the time-stamped data of their
442~live-day run~\cite{Aalseth:2011wp} for public use. This makes the
computation of the LS diagram in principle straight-forward. However,
the data also suffers from background activity of electron capture
decays of cosmogenically activated long-lived isotopes. A measurement
of higher energetic K-shell captures together with reported ratios of
L- to K-shell decays allows one to correct for the L-shell activity in
the DM acceptance region.

Based on the expected cosmogenic activity in the CoGeNT dataset it
seems reasonable to divide the low energy data into three regions:
$0.5-0.9\keVee$, $0.9-1.6\keVee$, and $1.6-3.0\keVee$. Considerable
cosmogenic activity is observed in the middle region. In contrast only
a small amount of cosmogenic activity is expected in the low region
($0.5-0.9\keVee$) and very little if any is expected in the high
region ($1.6-3.0\keVee$). Since the participating isotopes are rather
long-lived ($t_{1/2}\gtrsim 200\,\mathrm{days}$) they are expected to
result in substantial power at long periods in the corresponding
Fourier power spectrum. We have verified that this is indeed the
case. However, since we are interested in more localized phenomena in
frequency space, such as annual modulations, we need to subtract the
cosmogenic activity from the data. We have done this on a daily basis
based on the reported activity levels~\cite{Juan}. We have verified
that the results remain qualitatively the same even when adopting a
different subtraction procedure, where we fit for the exponentially
decaying component.

\begin{figure}[ht]
\begin{center}
\includegraphics[width=\columnwidth]{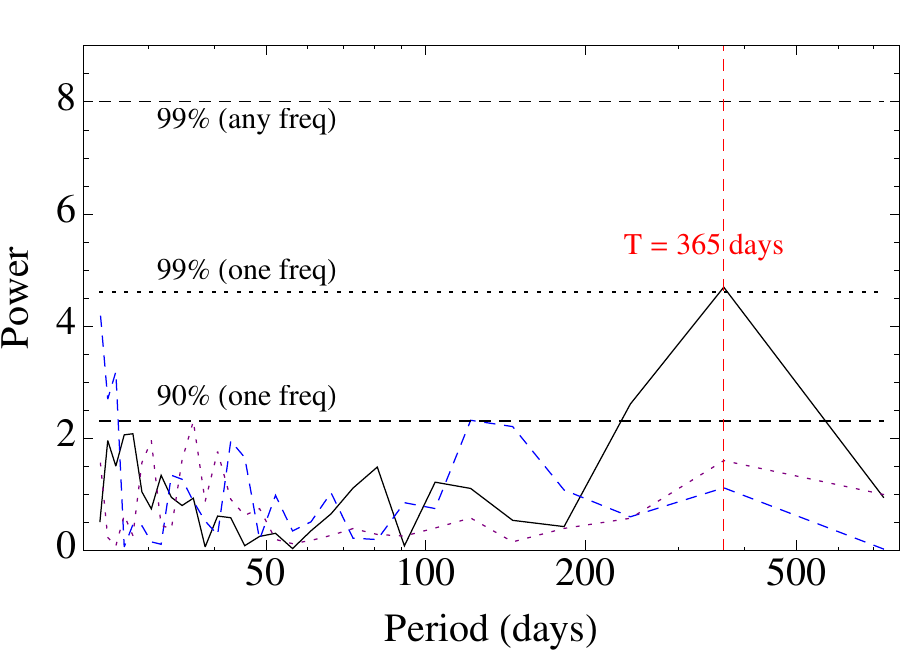}
\end{center}
\caption{The Lomb-Scargle power spectrum of the subtracted CoGeNT data
  as a function of the period for the three energy regions:
  $1.6-3.0\keVee$ (solid-black), $0.9-1.6\keVee$ (dashed-blue), and
  $0.5-0.9\keVee$ (dotted-purple).}
\label{fig:CoGeNT_null}
\end{figure}

Figure~\ref{fig:CoGeNT_null} shows the power spectrum for the subtracted
data in the three energy regions. The power spectrum was calculated in
30 equally spaced frequencies with a fundamental frequency of 2
years. The null hypothesis of noise  can only be confidently rejected
for the $1.6-3.0\keVee$ region, where considerable power is present at
a period of about one year. However, the same cannot be said about the
lower energy regions where no significant power is observed. Thus, one
can claim the detection of annual modulations at about $99\%$
confidence level only in the $1.6-3.0\keVee$ region.

\section{Phase Analysis}
\label{sec:phase}
Aside from the period, the second most important characteristic of the
oscillations observed by the DAMA experiment is the phase of the
signal. Dark matter collision rate with the detector is expected to
peak on June 2$^{\rm nd}$, corresponding to $t_0 =152$ days after
January 1$^{\rm st}$. In this section we investigate the phase
associated with the oscillations seen in the DAMA and CoGeNT data and
compare them to the phase of the muon intensity modulations. This
comparison was already attempted by the DAMA collaboration itself,
however, it was criticized by refs.~\cite{nygren,Blum:2011jf} on two
accounts. First, in their fit to the data the DAMA collaboration fix
the period and allow only the phase to float. Second, the underlying
signal may not be truly sinusoidal which may invalidate the
statistical inference of a fit to a sine function. We address the
first issue in this section and investigate the second problem in the
next.

\subsection{Frequentist Approach}

\begin{figure}[tb]
\begin{center}
\includegraphics[width=\columnwidth]{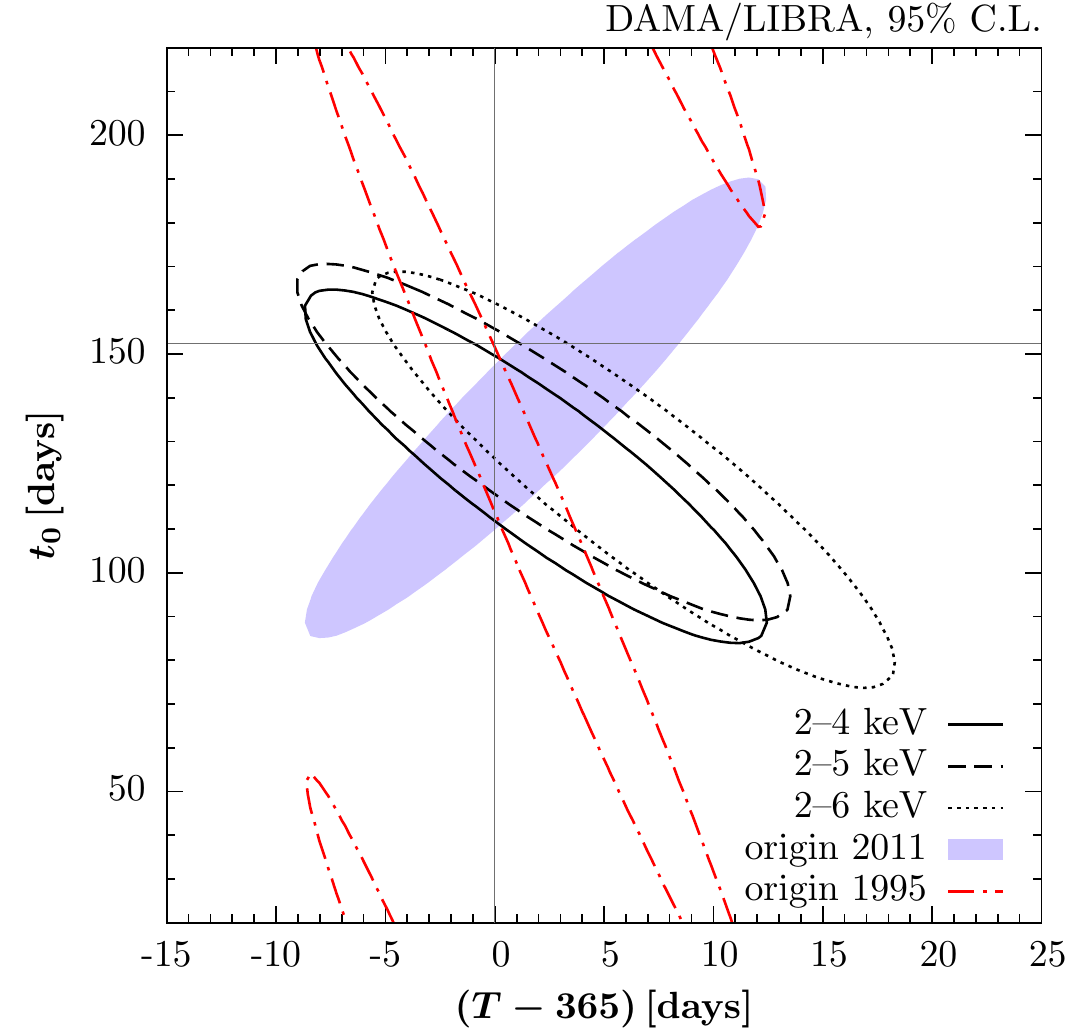}
\end{center}
\caption{Confidence regions in period $T$ and phase $t_0$ obtained
  from a $\chi^2$ fit to the DAMA/LIBRA residuals. The various lines
  as labeled correspond to the three different energy binnings
  provided by the DAMA collaboration with the time origin set to Jan
  1, 2003. The shaded region illustrates the effect of shifting the
  time origin to Jan 1, 2011 for the 2--4\,keV residuals and the light
  dot-dashed ellipse shows the time origin shift to Jan 1, 1995.}
\label{fig:freq}
\end{figure}

We start with a simple frequentist approach and investigate the effect
of departing from strict annual periodicity (i.e $T$ is
not necessarily 365 days) on the DAMA phase $t_0$. Under the premise
of a sinusoidal signal, $A\times\cos[\omega(t-t_0)]$ with
$\omega=2\pi/T$, we fit amplitude $A$ to the DAMA/LIBRA residuals by
minimizing the usual $\chi^2$ function while scanning over period $T$
and phase $t_0$. A confidence region in $t_0$ and $T$ can be
constructed from the profile likelihood method~\cite{pdg2010} which
effectively removes the dependence on the nuisance parameter $A$. This
way, we first obtain the global minimum $\chi^2_{\mathrm{min}} $ from
which the confidence region is obtained by requiring
\begin{equation}
  \label{eq:conf-reg}
  \chi^2(\hat A,t_0,T) \leq \chi^2_{\mathrm{min}} + \Delta \chi^2 ,
\end{equation}
where $\hat A$ is the maximum likelihood estimate for $A$ at each
point $(t_0,T)$.  For a coverage probability of 95\% one chooses
$\Delta \chi^2 = 5.99$.  Figure~\ref{fig:freq} shows the result of
fitting the DAMA/LIBRA residuals in the $(T,t_0)$-plane. The three
ellipses give the 95\%~C.L. regions for the various energy binnings as
provided by the DAMA collaboration. As can be seen the data---at the
required confidence---are not necessarily consistent with the dark
matter interpretation ($t_0=152$\,days and $T=1\,$yr as indicated by
the thin gray lines). As we argue in the next paragraph, the
interpretation of such fits have to be handled with some care.

When allowing the period to float it is important to realize that
statements about the inferred phase become dependent on the starting
date. First, there is the obvious effect that the phase is measured in
days with respect to Jan~1$^{\rm st}$, but if the period is not one
full year, then which year is used as the origin affects the
phase. Comparison of the phase in days between experiments with
different origins is meaningless until the origins are made to
coincide. This effect is simple to correct for and requires that we
use the same time origin for the different data sets. In
Fig.~\ref{fig:freq} the chosen time-origin for the three ellipses
showing the various energy bins is Jan 1, 2003.

There is another, more subtle issue that affects the determination of
the phase when the period is allowed to float. As can be seen from the
ellipses in Fig.~\ref{fig:freq} the DAMA data exhibits an
anti-correlation between the period and the phase. Fits with periods
longer than a year strongly favour phases smaller than $t_0\sim 150$
days and vice-versa. Note, however, that the sign of the correlation
itself depends on the chosen origin. When shifting the latter forward
such that the DAMA/LIBRA data lies in the past, $t_0$ and $T$ become
positively correlated instead. This is illustrated in
Fig.~\ref{fig:freq} by the filled ellipse obtained from the 2--4\,keV
energy bins for which we have chosen Jan 1, 2011 as the origin.
It is important to note that the distribution in $t_0$,
given a certain period, depends on the chosen time origin.  Ideally,
we would like to make coordinate-independent statements and care must
be taken when interpreting the results. This issue further motivates
the correlation study provided in the next section.

Given the above, before proceeding and comparing the LVD and DAMA data sets we must decide on a common time origin. The starting date used by the
DAMA collaboration (Jan 1$^{\rm st}$, 1995) is six years apart from
the LVD data (Jan 1$^{\rm st}$, 2001). In our analysis we will
concentrate on the DAMA/LIBRA data, which started on September 9$^{\rm
  th}$, 2003, and so a sensible choice is January 1$^{\rm st}$, 2003
as the origin since it has sufficient overlap with both data sets.
To quantify the level of agreement of the muon-induced background
hypothesis with DAMA we switch now to a Bayesian approach.  This allows for a convenient generalization of the Lomb-Scargle periodogram into the two dimensional phase-frequency space as discussed in Appendix~\ref{sec:bayes-vers-lomb} . Such an
approach is particularly convenient when one wishes to incorporate
further assumptions on the provided data sets (e.g. choosing priors on
the period, phase, and amplitude). We will not incorporate any such
assumptions in the current analysis, since we would like to keep it as
unbiased as possible. Moreover, we have verified that the conclusions
arrived at below remain the same when we employ a frequentist analysis
instead.

\subsection{Bayesian Approach}

The Bayesian approach can conveniently generalize the Lomb-Scargle periodogram and allow for the testing of different hypotheses and priors. In this section we
restrict ourselves to modelling the data with a simple sinusoidal
function $f(t)=A\times \cos{[\omega(t-t_0)]}$ but following through
the steps presented in Appendix~\ref{sec:bayes-vers-lomb}, a
generalization to more complicated test-functions is in principle
straightforward.

We are mainly interested in the relationship between phase $t_0$ and
period $T$. The posterior probability $P(\{\omega,t_0 \}||d)$,
\textit{i.e.} the probability of observing frequency $\omega =
2\pi/T$, and phase $ t_0$ given the data $d$ is obtained by
integrating the full posterior $P(\{A,\omega,t_0 \}||d)$ over the
amplitude $A$ restricted to positive values\footnote{It is possible to
  obtain a more compact expression for the posterior without the error
  function by integrating over all amplitude values both positive and
  negative. However, that leads to a $\pm\pi$ ambiguity in the
  phase. While this degeneracy is easily removable by inspection, we
  prefer to avoid this complication in what follows.},
\be
P(\{\omega,t_0 \}||d) &\propto& \frac{\sigma^{1-N}}{\sqrt{p}} \left[1+{\rm Erf}\left(\frac{h}{\sqrt{2\sigma^2 p}} \right) \right] \nonumber \\ &\times&  \exp\left(\frac{h^2}{2 \sigma^2 p}\right) ,
\label{eq:posterior}
\ee 
where,
\be
p&=& \sum_i \cos^2\omega\left(t_i-t_0\right) ,\\
h&=& \sum_i d_i \cos\omega\left(t_i-t_0\right) .
\ee

Before proceeding and using~(\ref{eq:posterior}) in the full
two-dimensional period-phase space we can make contact with the
procedure which is usually employed, \textit{i.e.} fixing the period
to one year, $T=365$~days. Thereby we impose a delta function prior on the
period centered at one year from which we consequently obtain the
posterior probability in phase only, $P(\{t_0 \}||d)$. We find that
the DAMA/LIBRA and LVD data do not agree with respective
values
\begin{align}
  \label{eq:phase-posterior}
t_0(\mathrm{DAMA}) & = (131\pm 13)\,\mathrm{days} ,\\ 
t_0(\mathrm{LVD}) & = (187\pm 2)~\mathrm{days}  .
\end{align}
The tight error bars on the LVD data may at first seem surprising. To
better motivate this number we consider that given the 8 years of
data, one can easily determine by eye a phase shift of about 20
days. Statistics allows a further reduction of about an order of
magnitude because of the very large number of points available in the
dataset. However, it should be kept in mind that the uncertainty
quoted is only statistical and does not take into account the possible
systematic issues associated with the obvious presence of temporal
variations on larger time scales\footnote{Moreover, there is a systematic error associated with the digitization of the data that amounts to an additional uncertainty of a couple of days.}. In fact, using the error bars of the
LVD data, the resulting $\chi^2$ of a fit to a sinusoidal function is
extremely poor. Hence, the numbers quoted above should really only be
understood as the best inference on the underlying parameters of a
model consisting of a single frequency.

\begin{figure}[ht]
\begin{center}
\includegraphics[width=\columnwidth]{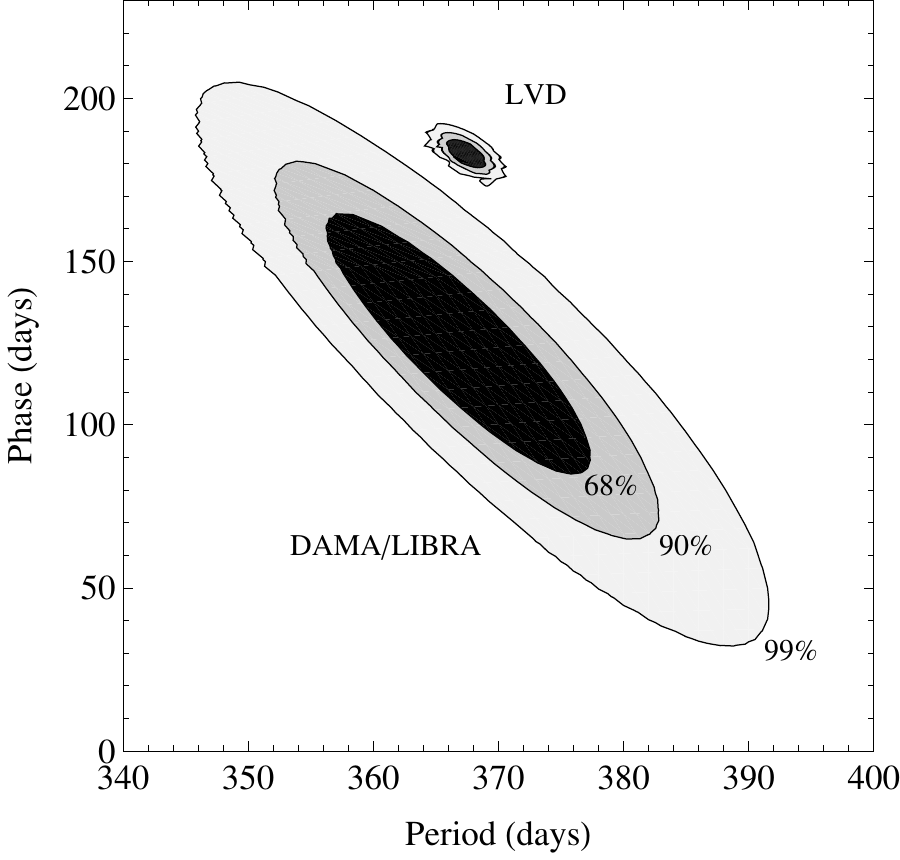}
\end{center}
\caption{A comparison of the period-phase posterior for DAMA/LIBRA
  (bottom) and the LVD phase (top). Shown are the 68\%, 90\%, and 99\%
  credibility ellipses.}
\label{fig:PeriodPhaseComparison}
\end{figure}

To see if the discrepancy between DAMA/LIBRA and LVD remains when
relaxing the assumption of a strict annual periodicity, we now
evaluate~(\ref{eq:posterior}) for the two data sets. The result is
shown in Fig.~\ref{fig:PeriodPhaseComparison}. This figure clearly
shows that the inferred modulations seen in the two data sets---if
interpreted as sinusoidal modulations---are incompatible with each
other.

\begin{figure}[ht]
\begin{center}
\includegraphics[width=\columnwidth]{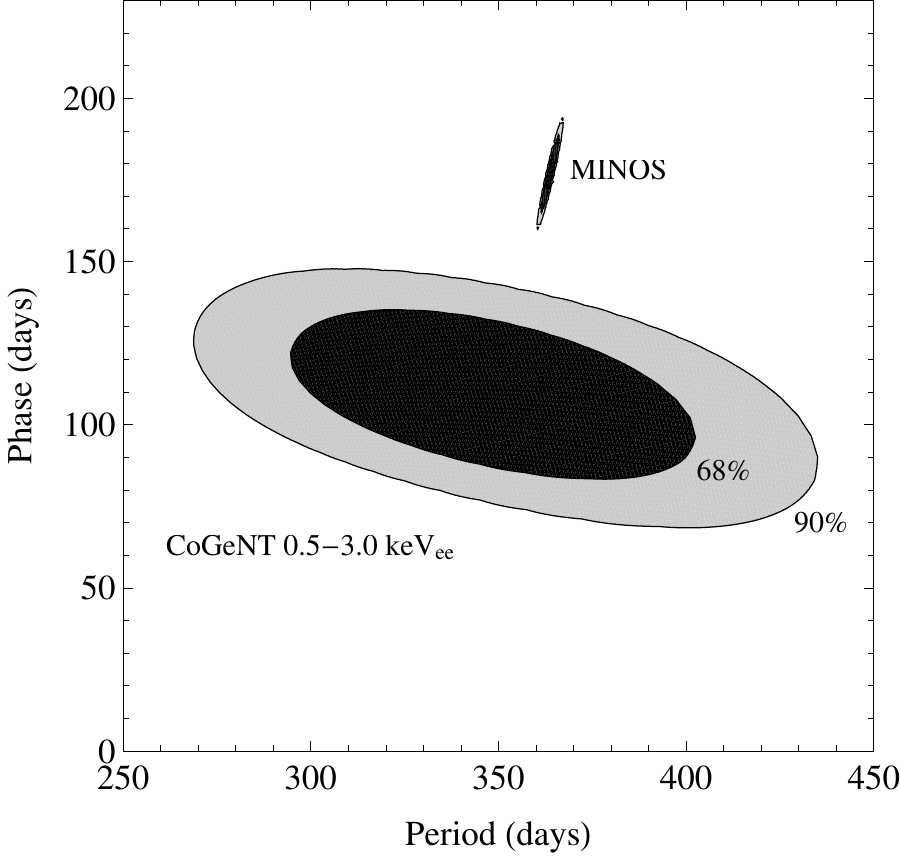}
\end{center}
\caption{The period-phase posterior for the full low energy CoGeNT
  data ($0.5-3.0\keVee$)~\cite{Aalseth:2011wp} and MINOS
  data~\cite{Adamson:2009zf}. Shown are the 68\%, and 90\% credibility
  ellipses. The time origin in both cases was chosen to be January
  1$^\text{st}$, 2010. }
\label{fig:CoGeNTPeriodPhase}
\end{figure}

A similar analysis can be done for the CoGeNT data%
\footnote{A Bayesian approach to the CoGeNT data has very recently
  also been taken by~\cite{Arina:2011zh}.},
as shown in Fig.~\ref{fig:CoGeNTPeriodPhase}, where we chose to use
the entire low energy range ($0.5-3.0\keVee$). As discussed in the
previous section, evidence for modulation is only present in the
restricted range of $1.6-3.0\keVee$. However, isolated power on an
annual time scale is present in the entire range and it is therefore
not unreasonable to employ the full range when looking to make
inferences about the modulation parameters.  CoGeNT's credibility
ellipses are compared with those obtained from the MINOS
data~\cite{Adamson:2009zf}. From there, we arrive at similar
conclusions as before, namely, that the muon and direct detection data
sets seem incompatible We have verified that when considering the
restricted range $1.6-3.0\keVee$ in the CoGeNT data we arrive at the
same conclusions.  We note in passing that the orientation of the
MINOS ellipses---which are obtained with a time origin of Jan
$1^{\mathrm{st}}$, 2010---is similar to the one found in
Fig.~\ref{fig:freq} for DAMA once one shifts the time origin to the
future of the data set.

The credibility contours in Fig.~\ref{fig:CoGeNTPeriodPhase} indicate
a range of viable parameters which is almost twice as large as what
has been quoted in the CoGeNT release paper~\cite{Aalseth:2011wp}.
For a direct comparison we thus also perform a frequentist fit to a
cosine function (plus a constant.) For example, when using the Minuit
package~\cite{minuit} for the $\chi^2$-minimization, we obtain $T =350
\pm 26$ days, $t_0 = 110\pm 13$ days, and $A = 17\%\pm 4\%$ for a time
binning resulting in 21~bins. This is in good agreement
with~\cite{Aalseth:2011wp}. However, it should be stressed that the
quoted errors are obtained from a default value of $\Delta \chi^2=1$
which does not reflect the increased freedom of fitting more than one
parameter. Furthermore $\chi^2 $ does not grow very large with respect to
its rather small minimum value, $\chi^2_{\mathrm{min}}/d.o.f. =
6.8/17$ and even a fit to a constant rate yields a reasonable
$\chi^2_{\mathrm{min}}/d.o.f. = 26.9/20$. This can be traced back to
large error bars in conjunction with limited data. As a result of
this, contours with nominally larger confidence will rapidly open
up the parameter space. This, however, does not signal real
compatibility of the data sets because of the aforementioned reasons.

\section{Correlation Analysis}
\label{sec:correlation}
A valid concern with regard to the analysis above is the underlying
assumption of sinusoidal variation. The power-spectrum of the LVD
data, Fig.~\ref{fig:LVD_null}, makes it clear that power exists in
modes with periods larger as well as smaller than one year.  One may
then worry that the phase comparison discussed in the previous section
suffers from a systematic misinterpretation. We do not entirely
endorse this concern because it is difficult to understand how the
very prominent annual modulation in the LVD data, with its very
definite phase, can be masked by the much weaker modulations at other
frequencies. Nevertheless, we now set to investigate this issue in a
way that does not rely on assuming any particular functional form for
the underlying temporal variations.

Moreover, even if the timing between muons and DAMA (and possibly
CoGeNT) are incommensurate at first sight, could it be that the
underlying stochastic nature of the background process alleviates the
observed tension in the annual phase? This explanation for DAMA was
suggested recently in an interesting paper by Blum~\cite{Blum:2011jf}.
Assuming a simple, generic model for how a muon-sourced background may
be realized, the distribution in phase originating from the Poissonian
process is claimed to be compatible with the one observed by DAMA.

In this section we address the above issues by performing a
correlation analysis. In particular we follow Ref.~\cite{Blum:2011jf}
and generate mock data for DAMA based on the muon hypothesis. We show
that the resulting mock data exhibits a substantially higher degree of
correlation with the actual muon data than does the real data from
DAMA thus ruling out the hypothesis. We then go on and use a similar
correlation analysis to show that the muon hypothesis is also an
unlikely cause for the modulations seen by CoGeNT.

\subsection{DAMA}
\label{sec:dama-corr}

Since there is considerable overlap between the DAMA/LIBRA and LVD
data, it is straightforward to define Pearson's coefficient of
correlation. We first bin the LVD data according to the DAMA/LIBRA
bins. Then the sample's correlation coefficient is defined as,
\be
\label{eq:pearsonr}
r_{X,Y} = \frac{1}{n-1}\frac{\sum_{i=1}^{n}\left(X_i - \bar{X}\right)\left(Y_i - \bar{Y}\right) }{\sigma_X\sigma_Y}
\ee
where $n$ is the number of overlap bins, $\bar{X}$ and $\bar{Y}$ are
the samples' mean, and $\sigma_{X,Y}$ are the samples' standard
deviations. The first statistical question we address is whether we
can exclude the no-correlation hypothesis. The answer to this question
lies in the statistic,
\be
t = \frac{r\sqrt{n-2}}{\sqrt{1-r^2}}
\ee
which---in the case of the null-hypothesis---is known to follow the
Student's distribution with $d.o.f = n-2$.  The DAMA/LIBRA and LVD
data share 39 temporal bins and so the 90\% (99\%) two-sided exclusion
limit on the statistic is $|t| > 1.687 (2.715)$. 
For the correlation between the two data sets we find a value,
\begin{equation}
  \label{eq:r}
   r_{\mathrm{LVD,DAMA}} = 0.44\quad \Rightarrow\quad t_{\mathrm{LVD,DAMA}} = 2.95  ,
\end{equation}
which allows us to exclude the no-correlation hypothesis with
confidence level greater than 99\%. This, however, should come as no
surprise since both datasets exhibit strong annual modulations with a
similar phase. Any such samples will exhibit a correlation at some
level, but that of course does not imply that they are indeed causally
related.  The more interesting question we would like to address now
is whether we can exclude causation. This question can only be
answered in a model-dependent way.

The model we consider was presented by Blum in
Ref.~\cite{Blum:2011jf}. It assumes that the DAMA count rate $s_i$ in
a time bin of width $\Delta t_i$ in an energy range $\Delta E$ for a
detector of mass $M$ is related to the muon flux $I_{\mu,i}$ during
that time by,
\be
\label{eqn:BlumModel}
s_i = \frac{y N_{\mu,i}}{M \Delta E \epsilon_i \Delta t_i},
\ee
where $N_{\mu,i}$ is Poisson distributed  with mean, 
\be
\langle N_{\mu,i} \rangle = A_{\rm eff} I_{\mu,i} \epsilon_i \Delta t_i.
\ee
Here $y$ is the number of signal counts/muon (yield), $A_{\rm eff}$ is
an effective area in which muons are collected, and $
\epsilon_i\approx 60\%$ is the duty cycle during time bin $t_i$. As
was argued in~\cite{Blum:2011jf}, direct hits of muons in the detector
would require $y\approx 50$ and would lead to a spread $\sigma_i/s_i$
in events which is about a factor of five larger than what is actually
observed. Since this is in clear conflict with the data, the other
possibility that we are going to consider is the secondary effect
muons can have due to the spallation reactions on nuclei and the
neutron flux they induce. In this case muons can be collected in a
larger area, say, $A_{\rm eff} \approx 10~{\rm m^2}$, which requires
an order one yield $y\approx 2$ only. Using the latter numbers we
generate $10^4$ realizations of DAMA data based on this model.

\begin{figure}[tb]
\begin{center}
\includegraphics[width=\columnwidth]{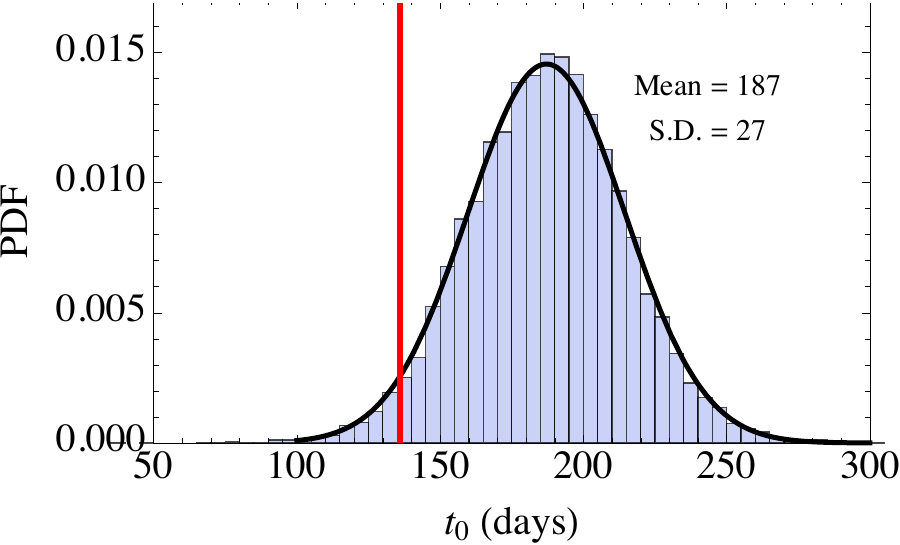}
\end{center}
\caption{The population distribution of the phase for the model by
  Blum, Eq.~(\ref{eqn:BlumModel}) compared with the best fit to data
  obtained by the DAMA collaboration, $t_0 = 136$~days, shown as the
  red vertical line. }
\label{fig:t0distro}
\end{figure}

We first attempt to make contact with~\cite{Blum:2011jf} by plotting
in Fig.~\ref{fig:t0distro} the distribution of $t_0$ obtained from a
$\chi^2$-fit to a sinusoidal function with floating period, phase, and
amplitude. There is little resemblance with the equivalent Fig.~3
presented in~\cite{Blum:2011jf}. The latter shows a very broad
distribution with substantial support between $90\lesssim t_0
/\mathrm{days}\lesssim 250$ and a peak at $t_0\approx
100\,\mathrm{days}$. We find from our Fig.~\ref{fig:t0distro} that
$t_0$ is normally distributed with a mean of $\langle t_0 \rangle =
187\,\mathrm{days}$ and $\sigma=27\,\mathrm{days}$. We believe that
the disagreement with Ref.~\cite{Blum:2011jf} is due to a different
choice of the time-origin. As we discussed in section~\ref{sec:phase}
once we allow the period to float, the phase $t_0$ loses its absolute
meaning and the allowed region in $t_0$ becomes sensitive to the
origin of time. By shifting the origin from Jan 1$^{\mathrm{st}}$ 2003
to the one DAMA uses when quoting their data, Jan 1$^{\mathrm{st}}$
1995, we find that the distribution in $t_0$ indeed broadens
significantly similar to the one found in~\cite{Blum:2011jf}.  Though
it may certainly be better to choose the time origin in 2003, the
discrepancy signals a more serious conflict: the distribution in $t_0$
does not provide us with a robust test statistic when assessing the
compatibility of the mock data with the observed phenomena.

\begin{figure}[t]
\begin{center}
\includegraphics[width=\columnwidth]{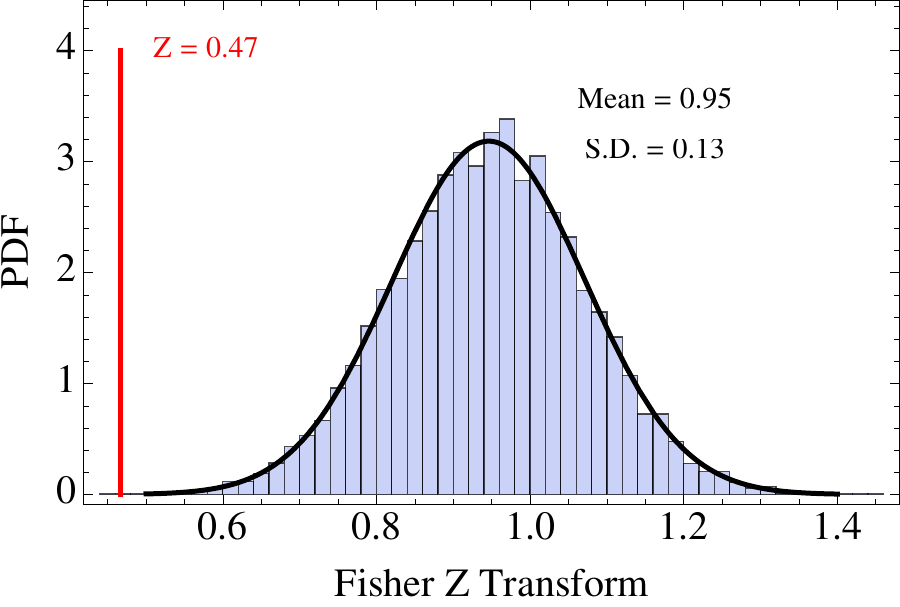}
\end{center}
\caption{The population distribution of the Fisher Z-transform for the model by Blum, Eq.~(\ref{eqn:BlumModel}) as compared to the sample value of $Z = 0.47$. Based on the population distribution, the model can be excluded with a confidence exceeding $99\%$. }
\label{fig:FisherZ}
\end{figure}

A better way to assess the (in)compatibility between the muon flux and
DAMA is to use the generated set of realizations and evaluate the
degree of correlation itself. This way we remain completely coordinate
independent in our statements and we can attempt to exclude the model
by showing that it implies a level of correlation much higher than
what is actually implied by the data. This is precisely what we now
labour to show. In general, if the model allows an exact calculation
of the population correlation coefficient $r_0$ then the Fisher Z
transform,
\be
Z = \tfrac{1}{2} \log\left(\frac{1+r}{1-r} \right)
\ee
is approximately normally distributed with mean and standard deviation
given by,
\be
\bar{Z} = \tfrac{1}{2} \log\left(\frac{1+r_0}{1-r_0} \right)\quad\quad\quad \sigma_Z = \frac{1}{\sqrt{n-3}}.
\ee
We will not use these results below since the model we consider does
not easily allow for an analytic evaluation of the population
correlation coefficient $r_0$. Instead we will utilize the numerical
realizations of the model and calculate the distribution
explicitly. As it turns out, however, the numerical results nicely
agree with the theoretical expectations for the shape of the
distribution.

Figure~\ref{fig:FisherZ} shows the results of the computation of the
Fisher Z-transform for each of these realizations. The distribution is
compared to the sample's Z-transform $Z = 0.47$ obtained from
(\ref{eq:r}). The numerical realizations reveal a much larger expected
correlation than what is observed in the data. Therefore the model is
excluded with a confidence greater than $99\%$.

\subsection{CoGeNT}
\label{sec:cogent}

We now move on to investigate the hypothesis that the reported CoGeNT
signal is caused by cosmic muons. In this respect, it is important to
note that the apparent modulation fraction of the CoGeNT signal is
most prominent in the high-energy bin $1.6-3.0\,\mathrm{keV_{ee}}$
with a value in excess of 10\%~\cite{Aalseth:2011wp,Fox:2011px}.
Seasonal muon-flux modulations of that order have indeed been reported
by IceCube~\cite{Tilav:2010hj}. However, the (somewhat) milder
climatic conditions at the Minnesota Soudan Mine location exhibit a
variation of at most $4\%$~\cite{Adamson:2009zf}. Therefore, unless
the scaling of the signal is---for some unknown reason---non-linear in
incident muon flux or unless the performance of the detector was not
stable throughout the data taking period, it is not possible to
generate, say, a 16\% CoGeNT modulation from a lesser modulated
sourcing process. This statement is independent of the potential
presence of further background. Even though this argument makes a muon
explanation of CoGeNT rather unlikely, we can still proceed and look
for a quantitative answer based on the temporal structure alone.

The main complication associated with such an analysis is the fact
that although measurements of the underground muon flux are available
from the MINOS experiment~\cite{Adamson:2009zf}, the data has no
temporal overlap with CoGeNT. We circumvent this issue by relying on
nearby atmospheric temperature data rather than on measurements of the
muon flux itself. As is well known, the underground muon flux is
tightly correlated with the (stratospheric) temperature. We therefore
choose to directly evaluate Eq.~(\ref{eq:pearsonr}) between the
effective atmospheric temperature parameter $T_{\mathrm{eff}}$ and the
background subtracted CoGeNT data. We refer the reader to
Appendix~\ref{sec:cosmic-muon-flux} for further details on this
procedure.

A correlation analysis is particularly appropriate in this case since
CoGeNT's relatively short data taking period (458 days) naturally
limits the significance of any statements about the annual periodicity
of the signal as evinced in Fig.~\ref{fig:CoGeNTPeriodPhase}. However,
the correlation analysis we employ requires binning the data, a
procedure that is complicated by the need to subtract the cosmogenic
background which is responsible for the majority of the event rate in
the $0.9-1.6\keVee$ region of the CoGeNT data. One may worry that such
subtractions introduce a bias depenedent on the choice of binning or
energy range.  Moreover, the unmodulated component of the subtracted
CoGeNT data exhibits a rise towards low energies that is not shared by
the modulated part, which if anything seems larger at higher
energies. Thus, it is not even clear which part of the energy spectrum
could be a result of the muon flux. In light of these complications we
consider various time- and energy binning in the hope of thoroughly
covering the different possibilities.

\begin{figure}[tb]
\begin{center}
\includegraphics[width=\columnwidth]{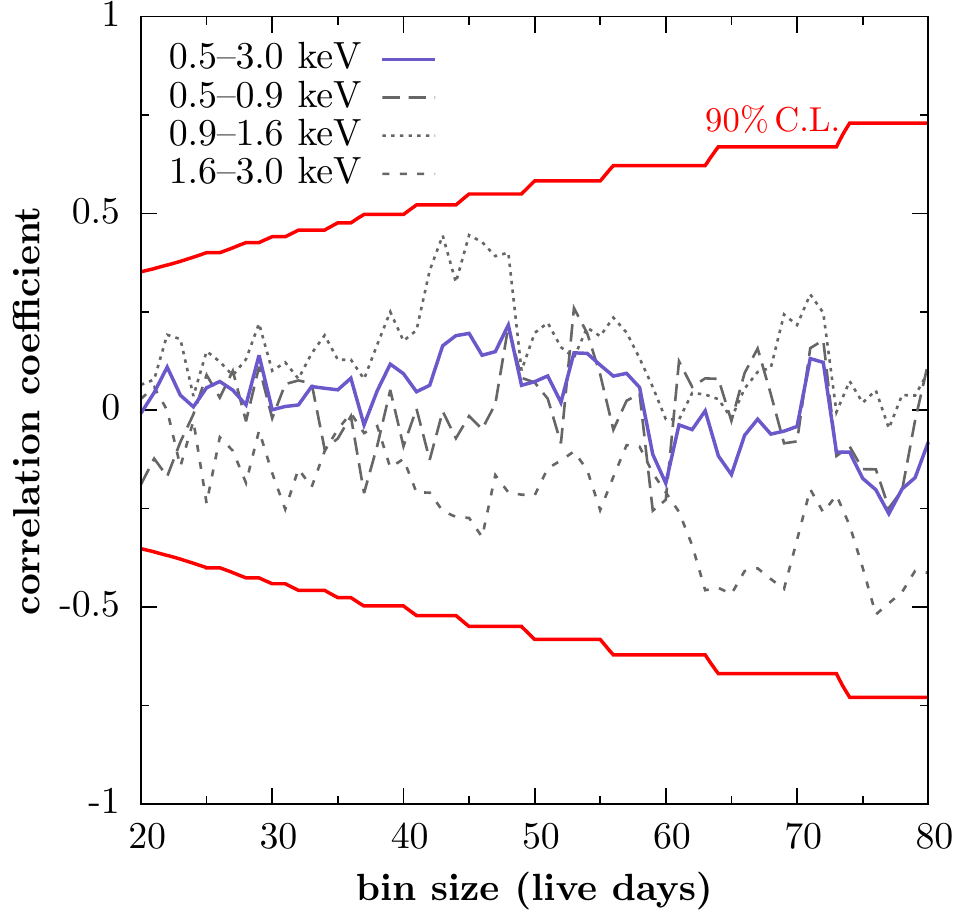}
\end{center}
\caption{Coefficent of correlation $r$ between $T_{\mathrm{eff}}$ and
  CoGeNT data as a function of bin size in live days. The various
  fluctuating lines correspond to different binnings in energy. None
  of the chosen energy regions show any significant degree of
  systematic, positive correlation.}
\label{fig:rcog}
\end{figure}

Figure~\ref{fig:rcog} shows the correlation coefficient as a function
of bin size in live days for various energy ranges. If any (positive)
correlation was present, which would corroborate the hypothesis that
muons are responsible for the observed signal, one would expect
significant degree of correlation independent of the bin size. With
the coefficient of correlation fluctuating near zero, the opposite is
observed. The step-like red lines delineate the (two-sided) 90\% CL
for rejecting the null-hypothesis. The data is therefore perfectly
consistent with the null hypothesis of no correlation. For larger bin
sizes $r$ begins to fluctuate more strongly. Although it is easier to
reach some degree of correlation with fewer bins it correspondingly
becomes harder to reach a given level of significance which is
indicated by the opening of the red lines.

\section{Biannual Modulations}
\label{sec:biannual}
As discussed throughout this work, one of the hallmarks of dark matter
direct detection is the annual modulation in the recoil rate, which is
the result of the Earth's motion with and against the WIMP wind
\cite{Drukier:1986tm, Freese}. However, as we show in this section,
one also expects higher harmonics (e.g. a period of half a year) to be
present at some level and those may prove useful as additional
confidence builders if a signal is detected.

\subsection{Harmonic Analysis}
The differential recoil rate, $dR/d\ER$ is a function of the Earth's
velocity relative to the rest frame of the dark matter halo, which
using the notation of Ref.~\cite{Savage:2009mk} is
approximately given by, \be
\label{eqn:vobs}
\mathbf{v}_{\rm obs} = \mathbf{v}_{\odot} + V_{\oplus}\left(\mathbf{\hat{\varepsilon}}_1 \cos \omega\left( t-t_1\right) + \mathbf{\hat{\varepsilon}}_2 \sin \omega\left(t-t_1 \right)\right)\quad \quad
\ee 
Here $\mathbf{v}_{\odot}$ is the velocity of the sun relative to the halo, and $\mathbf{\hat{\varepsilon}}_1$ ($\mathbf{\hat{\varepsilon}}_2$) is the velocity unit vector of the Earth's motion at $t_1 = $ March 21$^\text{st}$ ($t_1+\text{year}/4$).  
Due to Earth's orbit, the second term proportional to the Earth's relative velocity to the sun, $V_{\oplus} \sim 30 \text{\; km/s}$, modulates with frequency $\omega = \frac{2\pi}{1 \text{\; year}}.$  If the dark matter velocity distribution is isotropic, then the dark matter speed distribution in the earth's frame, $f(v)$, depends solely upon the magnitude of $\mathbf{v}_{\rm obs}$.  To a good approximation the magnitude of the velocity is given by 
\bea
|\mathbf{v}_{\rm obs}| \approx  |\mathbf{v}_{\odot}| + \frac{1}{2} V_{\oplus} \cos{\omega(t-t_0)}.
\label{eq:vobsmod}
\eea
Thus, dark matter scattering rate is a periodic function with a
fundamental period of 1 year, but like any periodic function it may
contain higher harmonic modulations of any frequency $\omega_n= n\,
\omega$. We therefore expect modulations in the scattering rate with
periods of $1/2, 1/3, \ldots$ a year corresponding to biannual,
triannual and higher frequency modulations.
In addition, there are harmonic corrections to $|\mathbf{v}_{\rm
  obs}|$ itself. They arise predominantly from corrections
to~(\ref{eqn:vobs}) due to smaller effects such as the ellipticity of
the earth's motion around the sun. Once these are taken into account,
they result in observable phase shifts between the different
harmonics.

We can learn more about the higher harmonics, by expanding in the parameter $\vratio = V_{\oplus}/2v_{\odot} \approx 0.06$.  
The time dependence of scattering arises solely due to the velocity dependence through 
\bea
\nonumber \frac{dR}{dE_R}\propto &&\int^\infty_{\vmin} \frac{f(v)}{v}dv\\ 
\label{eq:dRtime}
&\approx& \sum_{n=0,1,\ldots} \tilde{c}_n(\vmin) \left[\vratio \cos{\omega(t-t_0)}\right]^n \\ \nonumber
&=& 
\sum_{n=0,1,\ldots} c_n(\vmin) \cos{\left[n\omega(t-t_0)\right]}
\eea
Here we have used the fact that trigonometric identities allow powers of
$\cos{[\omega(t-t_0)]}$ to be re-expressed as sums of $\cos{\left[ n
    \omega(t-t_0)\right]}.$  In Eq.~(\ref{eq:dRtime}), $\vmin$ is the
minimum dark matter speed in the lab which can deposit a recoil energy
$E_R$,
\bea
\label{eqn:vmin}
\vmin = \frac{1}{\sqrt{2m_N E_R}} \left(\frac{m_N E_R}{\mu_{N\chi}} +\delta \right)
\eea
for a nuclear target of mass $m_N$ and dark matter-nuclei reduced mass $\mu_{N\chi} = m_N m_\chi/(m_N+m_\chi)$; in~(\ref{eqn:vmin}) we have included the possibility of inelastic scattering with a splitting $\delta$ in the dark matter sector \cite{iDM,MiDM}.   
The computation of the harmonic coefficients requires a choice of velocity distribution, and in what follows we consider the distribution proposed in~\cite{Lisanti:2010qx},
\be
f_k(v) \propto \left[ \exp\left(\frac{v_{\rm esc}^2 - v^2}{kv_0^2} \right) - 1\right]^k\Theta\left(v_{\rm esc} - v\right)
\ee 
Here, $\Theta(x)$ is the Heaviside step function, and we consider the dispersion $v_0 = 220 \text{\; km/s}$~\cite{Bovy:2009dr,McMillan:2009yr,Reid:2009nj} and the escape velocity $v_{esc} = 600 \text{\; km/s}$~\cite{Smith:2006ym}. 

\begin{figure}[tb]
\begin{center}
\includegraphics[width=0.96\columnwidth]{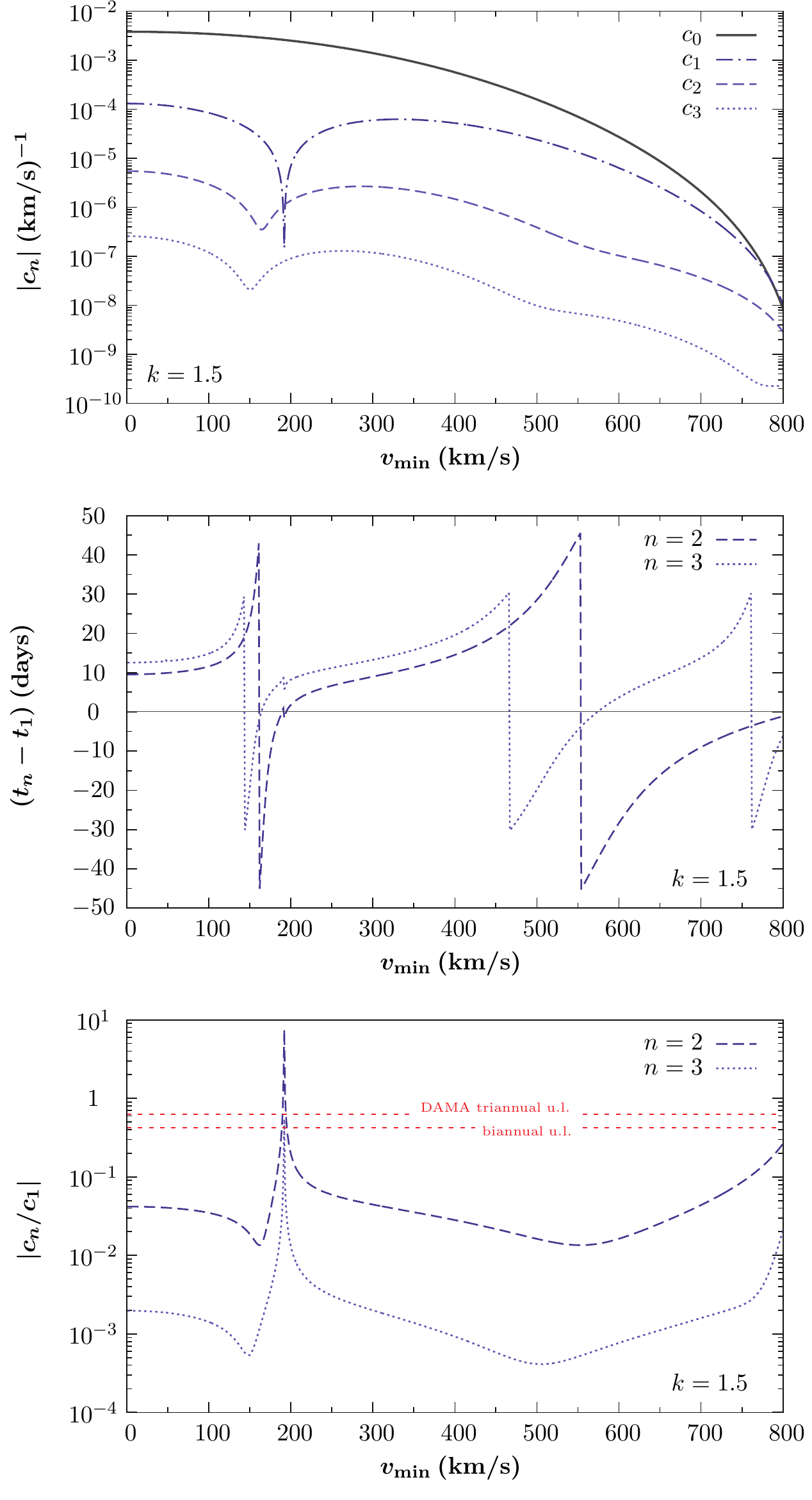}
\end{center}
\caption{The velocity dependence of dark matter scattering as a
  function of $\vmin$ for the $k$=1.5 velocity distribution proposed
  in \cite{Lisanti:2010qx}, with dispersion $v_0 = 220 \text{\; km/s}$
  and escape velocity $v_{esc} = 600 \text{\; km/s}$. The top plot
  shows the $|c_{n}|$ values for n=$0,1,2,3$ (top to bottom) where
  $c_n$ is in units of (km/s)$^{-1}$, the middle plot is the phase
  shift of the higher harmonics relative to the annual modulation,
  while the bottom plot shows the modulation amplitudes
  $|c_{n}/c_{1}|$ for n=$2,3$ (top to bottom). The horizontal lines in
  the bottom panel show the 90\% C.L. upper limits on a bi- and
  triannual signal in the DAMA data.}
\label{fig:amplitudes}
\end{figure}

In Fig.~\ref{fig:amplitudes} (top), we illustrate the behavior of the
harmonic coefficients, $c_n$, for the $k$=1.5 velocity distribution.
In order to plot the amplitudes on a logarithmic scale, we have taken
the absolute values of $c_n$.  The troughs in the plots indicate when
$c_n$ is changing sign. At high enough minimum velocity, $\vmin$,
the harmonic coefficients are all positive.  Interestingly, there
are~$n$ sign changes for~$c_n$, which can be understood from the behavior of
the velocity distribution~$f(v)$.

In the approximation of Eq.~(\ref{eq:dRtime}), the phases of all the
higher harmonics are the same as the annual modulation. However, as
mentioned above, there are higher order corrections to the magnitude
of the observer's velocity, Eq.~(\ref{eq:vobsmod}), that allow the
phases of the higher harmonics to deviate from the phase of the annual
modulation. Indeed, using an accurate parameterization of the earth's
velocity~\cite{Lewin:1995rx}, we find appreciable temporal shifts of
the biannual and triannual mode with respect to the annual one. This
is shown in the middle panel of~Fig.~\ref{fig:amplitudes}. The abrupt
changes seen in the figure indicate sign changes in the harmonic
coefficients $c_n$, which can be thought of as phase shifts
by~$1~\rm{yr}/2n$. The ambiguities in the phase shifts are fixed by
choosing the values which are closest to the annual one.  These phase
shifts serve as an additional signature which can be used to
discriminate a potentially positive signal from other sources.

In the bottom plot of Fig.~\ref{fig:amplitudes}, we plot the
modulation amplitude ratio $c_n/c_1$ for the biannual and triannual
modulation divided by the annual modulation.  There are two regions
where the higher harmonics are as important as the annual modulation.
The first region occurs near 200 km/s around the zero in the annual
modulation coefficient~$c_1$.  Here, the biannual amplitude~$c_2/c_0$
is at the per mille level which is small but potentially observable
with a large amount of exposure.  The second region where higher
harmonics are important is at high velocities, above the escape
velocity cutoff.  At these high velocities, there are substantially
less dark matter particles in the winter than in the summer.  Thus,
the scattering rate can vanish entirely in winter, with a chopped
cosine behavior, which requires more harmonics to fully describe the
approach to zero.  Although this would be obvious in a experiment with
no background, the presence of background can mask these higher
harmonics by eye, so it is useful to look for them.
 
We have also looked at the behavior of these modulation amplitudes for
the Standard Halo Model and also for a $k=3.5$ distribution
\cite{Lisanti:2010qx}.  The behavior at low $\vmin$ is very similar,
since that region depends mostly on the velocity dispersion and not on
the behavior near the tail.  The high velocity behavior enhances
(suppresses) the higher harmonics for sharper (shallower) escape
velocity cutoffs as seen in the $k=3.5$ distribution (Standard Halo
Model).
 
A detailed analysis of the detectability of these higher harmonics and
their ability to be faked by backgrounds is beyond the scope of this
paper.  Here, we restrict ourselves to a few remarks.  Any background
that modulates annually, with a small modulation amplitude will,
through the same Taylor expansion argument, also have higher harmonics
(although not necessarily all with similar phase).  What is peculiar
about dark matter is that the higher harmonic amplitudes are enhanced
at high $\vmin$ due to the escape velocity physics.  This could be
mimicked by background only if there was some effect during winter
which severely suppressed the background appearing as a nuclear
scattering event.

On the theory side, models with inelastic dark matter scattering lead
to enhanced higher harmonic amplitudes as compared with standard
elastic dark matter. A splitting $\delta \sim 100$~keV leads to values
of $\vmin$ which can be near the cutoff (see Eq.~\ref{eqn:vmin}).
In fact, we find that the biannual amplitude can be as large as 30\%
of the annual modulation amplitude for the 2-4 keV$_{ee}$ bins of
DAMA.  Looking at Fig.~\ref{fig:DAMA-LIBRA_null}, we see that this is
consistent with the amount of biannual power seen in these bins, since
the power ratio between biannual and annual modulation is $(c_2/c_1)^2
\lesssim 0.1.$ With a large increase in exposure, by a factor of
$O(10-100)$, DAMA would have enough statistics to begin testing this
and perhaps even see evidence for a nonzero biannual amplitude as we
show quantitatively in the next subsection.  

As a final remark, we note that it is possible to predict the energy
bin where the annual modulation should be suppressed, due to its
change of sign, which would be an interesting bin to look for biannual
modulation.  For heavy dark matter that scatters elastically, the
dependence on the dark matter mass drops out of the reduced mass, so
we can use Eq.~\ref{eqn:vmin}, to show that the energy bin $E_R = 90
\text{\; keV} (m_N/100 \text{\; GeV})$ is where the annual modulation
should be suppressed.  Such high energies might be hard to get substantial dark matter rates though, since the nuclear form factors tend to suppress such high energy recoils.  The situation gets better for dark matter much lighter than the nucleus, as the energy bin is $E_R = 0.9 \text{\; keV} (m_\chi/10 \text{\; GeV})^2 (m_N/100 \text{\; GeV})^{-1}.$  This requires knowledge of the dark matter mass to predict, but experimentally it might be verified that the sign of the annual modulation amplitude changes sign in a certain bin, thus it would be an interesting followup to see if there is a biannual mode in that bin.  In fact, the DAMA annual modulation amplitude is smaller in the first bin $2-2.5 \keVee$ bin, so it would be interesting to see if there is any biannual modulation there.

\subsection{Testing the Signal Hypothesis}

In the presence of a signal, the probability distribution function of the power $P$ in a particular frequency of the Lomb-Scargle power spectrum changes to~\cite{1975ApJS...29..285G,1982ApJ...263..835S},
\be
p(P; P_s) = I_0(2\sqrt{P ~P_s}) \exp\left(-P-P_s\right).
\ee
This formula assumes a normalized signal with power $P_s$ that is measured in units of the variance of the noise.
Given an observation of power $P_{\rm obs}$ we can reject a signal at a level of $1-\alpha$ by  requiring that, 
\be
{\rm Pr}(P>P_{\rm obs}) = 1 - \int_0^{P_{\rm obs}} p(P;P_s) dP = 1-\alpha
\ee
So, for example, from Figure \ref{fig:DAMA-LIBRA_null}, the DAMA/LIBRA
data in the $2-4\keVee$ energy range has a power of $P_{\rm obs}
=0.57$ (1.8) in the biannual (triannual) mode. This implies a
frequentist 90\% CL upper limit of $P_s^{\rm biannual} <1.9\, (4.4)$
on the power of the biannual (triannual) mode of a putative
signal. Since DAMA claims detection of an annually modulating signal
we can find the ratio of the higher harmonic power to the annual
power. The power observed in the annual mode of the same energy bin is
$11.0$. Therefore the fractional power in the biannual or triannual mode must be
lower than,
\begin{align}
P_s^{\rm biannual} &< 0.17 P_s^{\rm annual} , \nonumber \\ 
P_s^{\rm triannual} &< 0.4 P_s^{\rm annual}\quad \text{ at 90\% CL} .
\end{align}
Since the power is directly related to the square of the harmonic
coefficient, this limit implies a corresponding limit on the harmonic
coefficients, $c_2/c_1 < 0.42$ and $c_3/c_1 < 0.63$. These limits are
shown in the bottom panel of Fig.~\ref{fig:amplitudes}.
A similar analysis for the muon flux gives a 90\% CL upper limit on
the biannual mode of $P_s^{\rm biannual} < 6.3$, see Figure
\ref{fig:LVD_null}. Since the power observed in the annual cycle is
$P_s^{\rm annual} = 312$ this implies a limit on the harmonic
coefficients of $c_2/c_1 < 0.17$.

\section{Conclusions}
\label{sec:conclusions}
This paper was devoted to a detailed time-series analysis of Dark Matter direct detection data as well as related datasets.
The main findings are as follows: there is no evidence
to support the idea that atmospheric muons are responsible for the
signal that the DAMA collaboration observes; the annual modulations
observed in the CoGeNT data are statistically significant only in the
higher part of the low energy spectrum ($1.6-3.0\keVee$); no
correlation of the latter with the expected muon flux is observed;
biannual modulations may serve as an additional handle on any putative
dark matter signal.

With regards to the muon hypothesis discussed in refs.~\cite{ralston,
  nygren, Blum:2011jf} whereby atmospheric muons reaching the DAMA
detector are responsible for the signal observed, our analysis
indicates three significant difficulties with this idea:
\begin{enumerate}
\item The power spectrum of the LVD data on the muon flux indicates
  considerably more power at periods longer than one year compared
  with the DAMA/LIBRA data. This remains true even after we follow the
  undesirable procedure of the DAMA collaboration whereby the
  residuals are computed by subtracting the mean of the data on a
  yearly basis. This procedure is undesirable since it masks power at
  long periods. We therefore urge the DAMA collaboration to release
  the unsubtracted data to allow for a full comparison of the power
  spectrum. Similar conclusions have been reached
  in~\cite{Blum:2011jf}.
\item When fitting the two datasets to a sinusoidal function with
  variable amplitude, phase, and period the most likely regions for
  the two fits do not overlap.  On this point we are in disagreement
  with ref.~\cite{Blum:2011jf} which finds that the phase and period
  of the muon dataset can be in agreement with the DAMA signal. We
  believe that the source of this discrepancy is the choice of time
  origin which becomes important once the period is allowed to float.
\item A correlation analysis which is independent of any choice of
  functional form or fit to the datasets excludes the simple, yet very
  general model presented by ref.~\cite{Blum:2011jf} for connecting
  the muon flux with the DAMA signal.  It indicates that such a model
  would imply a much stronger degree of correlation than what is in
  fact observed in the data.
  Thereby, this analysis also puts to question the
  hypothesis~\cite{ralston,nygren} itself that the variation of the
  muon flux has a casual connection with the one seen by DAMA.
\end{enumerate}

Aside from these three difficulties, which depend only on the temporal
structure of the two signals, there also seems to be some difficulty
reconciling the amplitude of the muon modulations with that of the
DAMA signal.  On the face of it, the amplitude of the modulations
observed by LVD seems to agree well with the $\sim2\%$ modulations as
originally quoted by the DAMA collaboration. However, the full
single-hit energy spectrum as measured in
DAMA-LIBRA~\cite{Bernabei:2008yi}, seems to suffer from $^{40}$K
contamination in the signal region. Moreover, it was recently pointed
out in ref.~\cite{Kudryavtsev:2010zza} that the radioactive background
rates in the DAMA experiment are likely larger than what was estimated
by DAMA. If true, that would imply a smaller unmodulated amplitude of
the putative signal and hence a larger modulation fraction. Thus, any
model attempting to explain the DAMA signal as a result of the muon
flux will require some sort of nonlinearity in the signal yield.

Given all of the above, we find the muon hypothesis in its current
form rather unpersuasive. Any further investigations into this
possibility should address the aforementioned difficulties.

With regards to the CoGeNT data our analysis concentrated on three
different energy regions, $0.5-0.9\keVee$, $0.9-1.6\keVee$, and
$1.6-3.0\keVee$. We find that the presence of annual modulations can
only be firmly established in the $1.6-3.0\keVee$ region. The other
two regions require a somewhat more careful treatment of the data to
remove the contribution from the cosmogenic activity which contributes
to the power spectrum at long periods due to the long-lived isotopes
involved. Nevertheless, none of the procedures we have followed for
subtracting the cosmogenic contributions resulted in a statistically
significant yearly modulation signal in the subtracted data.  The
phase analysis of the modulations shows them to peak around the
middle of April, which is somewhat earlier than expected from DM recoils, but the uncertainty on this number is still rather large. Indeed, we find that the confidence intervals in the full two dimensional phase-period space are much too large for any significant statement to be made about the modulations observed. 

As in the case of DAMA we find no significant correlation between the
CoGeNT data set and the yearly modulation of the muon flux. In the absence of published
measurements of the latter for the time-period of the CoGeNT run we
used the atmospheric effective temperature inferred from
climate data instead. The one-to-one relation between the variation in
temperature and muon-flux allows us again to indirectly disfavor the
muon hypothesis as it applies to the modulations seen by CoGeNT. 

One can also wonder if muon induced backgrounds instead have a delayed signal, would that allow them to explain these dark matter anomalies?  As shown in Appendix~\ref{sec:activation}, production of long-lived radioactive isotopes does not result in any improvement.  The largest allowed temporal shift of the signal peak is to the beginning of October.  To complicate matters, the modulation amplitude of such a signal is always smaller than the muon amplitude.  These limitations make it difficult to explain DAMA or CoGeNT with such muon backgrounds.  

Finally, we discussed a new diagnostic for dark matter direct detection experiments.  Since the expected recoil rate of dark matter
against matter in the lab is a general periodic function with a
fundamental period of one year, it also contains biannual modulations
as well as higher harmonics.  The relative amplitude of the velocity
modulation is roughly $V_{\oplus}/2v_{\odot} \approx 0.06$, which is
rather small. The higher harmonics in the recoil rate have similar
phase to the annual modulation, but are generally suppressed.
However, we presented a few cases where they might be observed given
sufficient exposure. At the very least, the biannual mode can be
looked for and limits on its power can be obtained in a
straightforward fashion.

\begin{acknowledgments}
   We acknowledge helpful discussions with L.~Dai, M.~Kamionkowski, D.~Morrissey and
   M.~Pospelov. We thank J.~Collar for assistance in interpreting the
   CoGeNT data.
\end{acknowledgments}

\appendix
\renewcommand{\theequation}{A-\arabic{equation}}
\setcounter{equation}{0}

\section{Bayesian Version of the Lomb-Scargle Periodogram}
\label{sec:bayes-vers-lomb}

In this appendix we produce a derivation of the Lomb-Scargle
periodogram based on Bayesian analysis. This result was first derived
in Ref.~\cite{Bretthorst}. We present it here in order to introduce
the terminology, methodology, and notation used in deriving some of
the other results, as well as in order for this paper to be as
self-contained as possible.

We begin by considering the oscillatory model function~$f$ with a single harmonic frequency $\omega$,
\be
\label{eq:hypothesis}
f(t) = A \cos\left(\omega t\right) + B \sin\left(\omega t\right), 
\ee
as the hypothesis for the signal. The data points are therefore given by, 
\be
d_i = f(t_i) + n_i
\ee
where the data is taken to have zero mean $\langle d \rangle = 0$. The noise is denoted by $n_i$ which is assumed to be normally distributed with zero mean and variance $\sigma^2$. Under the hypothesis~(\ref{eq:hypothesis}) the probability $P(d||\{A,B, \omega \})$ of observing the data $d$ given the parameters $\{A,B, \omega \}$ is then directly proportional to the likelihood function 
\be
\nonumber
L(\{A,B,\omega\}, d) &=& \prod_{i=1}^N \frac{1}{\sqrt{2\pi\sigma^2}} \exp\left\{-\frac{[d_i-f(t_i)]^2}{2\sigma^2}\right\} \\ &=&  \left(\frac{1}{\sqrt{2\pi\sigma^2}}\right)^N \exp\left(-\frac{N Q}{2\sigma^2}\right) , 
\ee
where $N$ is the number of data points and
\be
\nonumber
Q &=& \langle d^2\rangle - \frac{2}{N} \left[A \sum_i d_i \cos\left(\omega t_i\right) + B  \sum_i d_i \sin\left(\omega t_i\right)\right] \\ \nonumber &+& \frac{1}{N}\left[ A^2  \sum_i \cos^2\left(\omega t_i\right) + B^2  \sum_i \sin^2\left(\omega t_i\right) \right. \\ &+& \left. 2AB  \sum_i \cos\left(\omega t_i\right)  \sin\left(\omega t_i\right)  \right] .
\ee
Bayes' theorem then allows us to obtain the posterior distribution $P\left(\{\omega,A,B \}|| d\right)$, \textit{i.e.} the probability of the parameters $\{\omega,A,B \}$ given the data $d$,
\be
P\left(\{\omega,A,B \}|| d\right) \propto P(\{\omega, A, b\}) L(\{A,B,\omega\}, d).   \quad
\ee

Assuming a flat prior on the nuisance parameters A and B, we can integrate over A and B to obtain the posterior distribution of $\omega$ alone. This is a gaussian integral, and would be trivial except for the cross term between A and B in $Q$. This term can be eliminated with the introduction of a phase-shift, $\tau$, by defining $\tilde{t}_i\equiv t_i -\tau$ and using the fact that $\langle d \rangle = 0$,
\be
\nonumber
Q &=& \langle d^2\rangle - \frac{2}{N} \left[A \sum_i d_i \cos\left(\omega \tilde{t}_i\right) + B  \sum_i d_i \sin\left(\omega \tilde{t}_i\right)\right] \\  &+& \frac{1}{N}\left[ A^2  \sum_i \cos^2\left(\omega \tilde{t}_i\right) + B^2  \sum_i \sin^2\left(\omega \tilde{t}_i\right)   \right] .
\ee
Effecting the gaussian integral over A and B gives,  
\be
\nonumber
P(\{\omega\} \| d_i) &\propto& \int dA~dB ~P(A,B) ~L(\{A,B,\omega\}, d)  \\ \nonumber &=&  \left(\frac{1}{\sqrt{2\pi\sigma^2}}\right)^{N-2}\frac{1}{\sqrt{\sum_i \cos^2\left(\omega \tilde{t}_i\right)\sum_i \sin^2\left(\omega \tilde{t}_i\right)}} \\&\times& \exp\left\{-\frac{1}{2\sigma^2}\left[N\langle d^2\rangle - {\rm LS}(\omega)\right] \right\} ,
\label{eqn:AppAposterior}
\ee
where the LS function is,
\be\nonumber
{\rm LS}(\omega) &=& \frac{1}{\sum_i \cos^2\left(\omega \tilde{t}_i\right)}\left[\sum_i d_i \cos\left(\omega \tilde{t}_i\right) \right]^2 \\ &+& \frac{1}{\sum_i \sin^2\left(\omega \tilde{t}_i\right)}\left[\sum_i d_i \sin\left(\omega \tilde{t}_i\right) \right]^2 .
\ee
The amplitude of the exponential in the posterior distribution, Eq.~(\ref{eqn:AppAposterior}), is a rather mild function of the frequency and so the most likely value for the frequency is the one which maximizes the LS function. This result is the Bayesian version of the usual frequentist LS statistics.

This approach lends itself to several generalizations and extensions. Ref.~\cite{Bretthorst} derives the posterior for the frequency associated with a sinusoidal signal with general coefficients (not necessarily constants). It is also possible to derive the posterior on the phase of a sinusoidal signal following similar steps to the above.

\renewcommand{\theequation}{B-\arabic{equation}}
\setcounter{equation}{0}

\section{Cosmic muon flux}
\label{sec:cosmic-muon-flux}

In this section we discuss the connection between cosmic ray induced
muon flux and its seasonal variation. Both, DAMA and CoGeNT, are
operated deeply underground ($>$~2000~mwe) so that these detectors are
well shielded from cosmic rays. Only muons with initial energies
$\mathcal{O}(\mathrm{TeV})$ are able to reach the respective LNGS and
Soudan Mine underground facilities. The highly energetic muons are
sourced from primary meson ($\pi$'s, $K$'s,\dots) decays which
themselves are produced when cosmic rays hit the atmosphere.
Since the chance for primary mesons to undergo further interactions
grows with air density, there is an increasing competition between
further degradation and decay when going to lower atmospheric levels.

The above chain of arguments leads to the following picture for the
high-energy part of the muon flux: the production proceeds
predominantly at stratospheric levels which are mainly subject to
seasonal temperature (and thus air density) variations. The
troposphere---with its daily modulation in temperature---only plays a
minor role in determining the muon flux. The phenomenological
relationship between muon flux and temperature can be written as
\begin{equation}
  \label{eq:muflux}
 \frac{\Delta I_{\mu}}{\langle I_{\mu} \rangle} = \alpha_T \frac{\Delta T_{\mathrm{eff}}}{\langle T_{\mathrm{eff}} \rangle} ,
\end{equation}
where $\langle I_{\mu}\rangle$ is the (average) integral underground
flux (with variation $\Delta I_{\mu}$) and $\alpha_T$ is an effective
temperature coefficient with inferred values $\alpha_T\simeq
0.93$~\cite{D'Angelo:2011fs} and $\alpha_T\simeq
0.87$~\cite{Adamson:2009zf} for the LNGS and Soudan Mine sites,
respectively.
The effective temperature $T_{\mathrm{eff}}$ is obtained from the
weighted average
\begin{equation}
  \label{eq:Teff}
   T_{\mathrm{eff}} = \frac{\int_0^{\infty} dX\, T(X) W(X)  }{ \int_0^{\infty} dX\,  W(X)  }
\end{equation}
where $X$ is the atmospheric depth. The weight function $W(X)$ depends
on parameters such as the amount of inclusive meson production in the
forward fragmentation region, atmospheric attenuation parameters of
mesons, and the muon spectral index. When computing $T_{\mathrm{eff}}
$ we use the expression for $W$ given in~\cite{Adamson:2009zf} which
includes the contribution of pions as well as kaons.

\begin{figure}[tb]
\begin{center}
\includegraphics[width=\columnwidth]{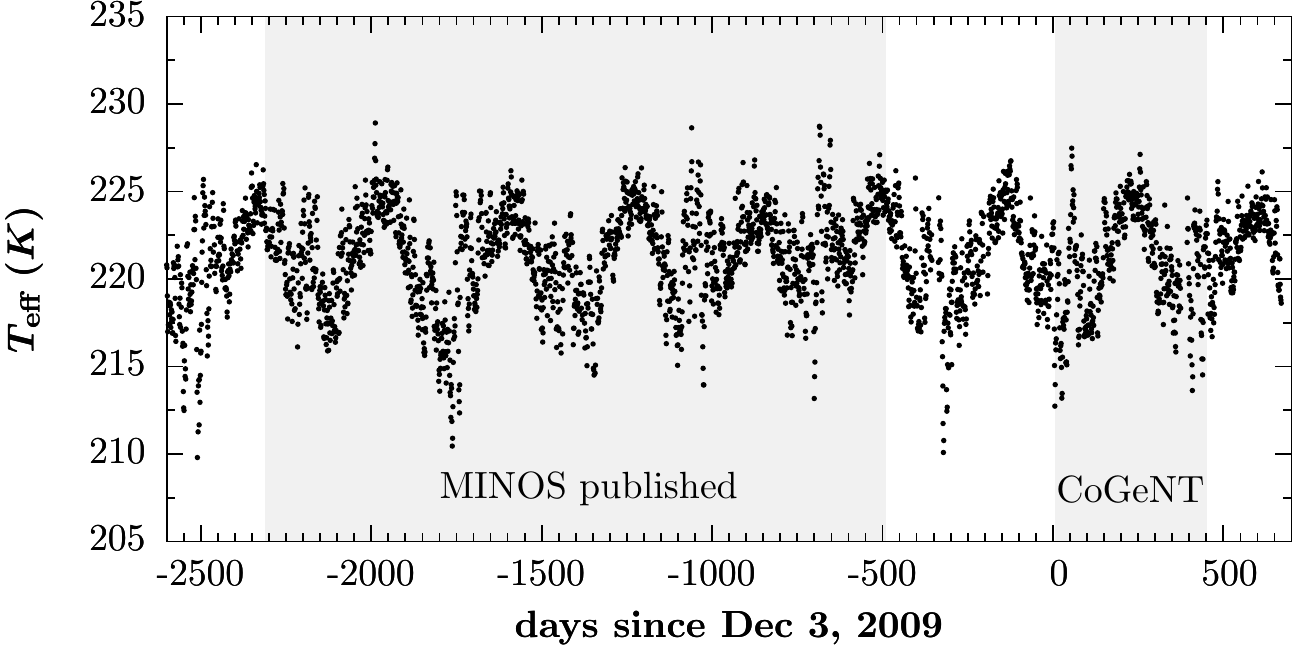}
\end{center}
\caption{Evolution of effective temperature parameter
  $T_{\mathrm{eff}}$ at the Soudan site as a function of time. The
  gray shaded regions indicate the overlaps with data-taking periods
  of MINOS and CoGeNT as labeled.}
\label{fig:teff}
\end{figure}

The tight correlation between $\Delta I_{\mu}$ and $\Delta
T_{\mathrm{eff}}$ is firmly established by a whole range of
independent measurements and at various geographic underground
locations, such as at
LNGS~\cite{Ambrosio:1997tc,Selvi,D'Angelo:2011fs}, at the Soudan
Mine~\cite{Adamson:2009zf}, or at the South Pole~\cite{Tilav:2010hj}.
For DAMA, the relevant LVD measurements are discussed in the
introduction. 
Turning to the CoGeNT site, the MINOS far detector has recently
reported measurements on the underground muon flux spanning a total of
five years~\cite{Adamson:2009zf}. Unfortunately, the published data
does not overlap with the CoGeNT run starting Dec 4, 2009 and lasting
458 days. The results are nevertheless relevant as they corroborate
the tight relation between effective temperature and measured muon
flux. In order to make statements about the muon background hypothesis
for CoGeNT we have obtained climate data from the Integrated Global
Radiosonde Archive (IGRA) for the nearby International Falls (MN)
weather station~\cite{durre2006overview}.  We compute daily
$T_{\mathrm{eff}}$ values by evaluating a discretized version
of~(\ref{eq:Teff}) including only high-quality data reaching
atmospheric pressure levels below 40~hPa. The result of this procedure
is shown in Fig.~\ref{fig:teff} which presents effective temperature
values for a time span covering both, MINOS and CoGeNT, as
indicated. We note in passing that the same IGRA data set was also
used by the MINOS collaboration itself to confirm $T_{\mathrm{eff}}$ values
obtained from another climate archive with good agreement.

\renewcommand{\theequation}{C-\arabic{equation}}
\setcounter{equation}{0}

\section{Activation of long-lived isotopes}
\label{sec:activation}
A loop-hole that exists in the analysis of assessing the viability of
the muon hypothesis for DAMA and CoGeNT is the following: What if one
significantly breaks the temporal coincidence between muon arrival
time in and around the detector and the time at which the signal is
produced. ``Significantly'' here means a shift in time which at least
on the order of weeks and more. It is clear then that one cannot
compare the time series of muon flux data with the one from DAMA and
CoGeNT at face value.

The muon flux itself can give rise to cosmogenically activated
isotopes. For example, the most inner shielding of the DAMA detector
is made from massive copper so that for example a reaction
${}^{65}\mathrm{Cu} + \mu^{+} \to {}^{65}\mathrm{Zn} + \bar\nu_{\mu}$
with a delayed decay of ${}^{65}\mathrm{Zn} $ with a half-life of
$244$~days is conceivable. Though the weak-scale cross section draws
doubt on the efficiency of such a process, neutron capture reactions
have cross sections many orders of magnitude larger. Indeed, it was
one of the central points in Ref.~\cite{ralston} to point to the
importance of $(n,\gamma)$ reactions for background considerations in
direct detection experiments. Let us at this point not worry about the
particular process but rather tackle the problem from the time series
point of view.
The question we are after is whether it is possible to induce a
time-delay $\hat t$ in the signal such that it features a maximum in
the year which is compatible with DAMA (and possibly by CoGenT.) As we
have seen, the muon flux peaks at least a month later than the signals
under consideration.  Therefore, one requires $\hat t \gtrsim
300$~days.

For simplicity, let us model the muon flux as a sinosoidal function,
$I_{\mu}(t) = I_0 [ 1 + I_m \cos{(\omega t)} ]$ where we are choosing
the origin of time to coincide with its maximum. Call $n$ the number
density of a stable element and $n^{*}$ its ``excited'' state formed
upon a generalized interaction with the muon flux and which decays
with a lifetime of $\tau = \lambda^{-1} $. The abundance
of $n^{*}$ is obtained by solving
\begin{equation}
  \label{eqn:rateeqn}
  \frac{dn^{*}( t)}{dt} = - \lambda n^{*}(t) + S_0\left[ 1 + S_m \cos{(\omega t)}  \right]
\end{equation}
with an overall source term $S_0$ with modulating fraction $S_m$ for
activation. In the physical picture, $S_0 = \lambda_{\mathrm{act}} n$,
where $\lambda_{\mathrm{act}}$ is the rate for activation and given by
the product of the activating (neutron) flux times the cross section
(with a potential average over the flux energy spectrum.) Clearly,
$S_0\propto I_0$ holds with the same phase and modulation amplitude in
Eq.~(\ref{eqn:rateeqn}) as in $I_{\mu}(t)$.

If the events seen by DAMA (or CoGeNT) are somehow related to the
decay of the radioactive isotope then the event rate in the detector
will be proportional to
%
$R \propto \lambda n^{*}(t) $
%
which is just the usual relation for the number of radioactive decays.
The rate equation Eq.~(\ref{eqn:rateeqn}) can be solved exactly,
\be n^{*}(t) = C_0 e^{-\lambda t} + S_0\tau + \tilde S_m \cos\left(\omega t -
  \theta\right) ,
\ee
where $\tan\theta = \omega/\lambda$ and $\tilde{S}_m = S_0 S_m/\sqrt{\lambda^2+\omega^2}$. The first term is transitory and
vanishes for $t\gg \tau$. Therefore, the event rate will approach
\be
R \sim S_0 + \lambda \tilde S_m\cos[ \omega\left( t - \hat t \right) ],
\ee
with the time delay, 
%
$\hat t = \theta / \omega $. 
%
The latter is always a positive quantity with a minimum of zero days
and maximum of $91.25$ days (quarter year). The phase delay is maximal
($\theta = \pi/2$) when the lifetime of the decay is much longer than
one year $\tau \gg 1\,\text{yr}/2\pi$.  Also, interestingly, the
modulation amplitude is always smaller than the muon flux, since
$\lambda \tilde{S}_m/S_0 = (\lambda/\sqrt{\lambda^2+\omega^2}) I_m.$
Thus, we see that this muon related background is not optimal for
explaining DAMA or CoGeNT, since the maximum of the signal can only be
shifted into the fall, but not further into spring as would be
required by DAMA and/or CoGeNT, and the modulation amplitude of the
signal is always reduced from the muon's amplitude.

\bibliography{biblio}
\end{document}